\documentclass[journal,onecolumn,10pt,twoside]{IEEEtran}
\newtheorem{proposition}{{Proposition}}
\newtheorem{definition}{{Definition}}
\newtheorem{theorem}{{Theorem}}
\newtheorem{lemma}{{Lemma}}

\usepackage{cite}
\usepackage{amsmath,amssymb,amsfonts}
\usepackage{algorithmic}
\usepackage{graphicx}
\usepackage{textcomp}
\usepackage{xcolor}

\normalsize

\ifCLASSINFOpdf
\else
\fi

\begin{document}
%
\title{Rate-Distortion-Perception Theory for the Quadratic Wasserstein Space}
%
%
%

\author{Xiqiang~Qu,   Jun~Chen, Lei~Yu,
	and~Xiangyu~Xu
}

\maketitle

\begin{abstract}
We establish a single-letter characterization of the fundamental distortion-rate-perception tradeoff with limited common randomness under the squared error distortion measure and the squared Wasserstein-2 perception measure. Moreover, it is shown that this single-letter characterization can be explicitly evaluated  for the Gaussian source. 
Various notions of universal representation are also clarified. 
\end{abstract}

\begin{IEEEkeywords}
Common randomness, Gaussian source, optimal transport, lossy source coding, MMSE estimate, rate-distortion-perception theory, soft-covering lemma, squared error, universal representation, Wasserstein distance, weak convergence.
\end{IEEEkeywords}

%
\IEEEpeerreviewmaketitle

\section{Introduction}

In lossy source coding, the encoder compresses an i.i.d. source sequence $\hat{X}^n:=\{X(t)\}_{t=1}^n$ into a bit string, while the decoder produces a reconstruction sequence $\hat{X}^n:=\{\hat{X}(t)\}_{t=1}^n$ based on this bit string. The end-to-end distortion is quantified as
\begin{align}
	\frac{1}{n}\sum\limits_{t=1}^n\mathbb{E}[d(X(t);\hat{X}(t))],
\end{align}
where $d:\mathcal{X}\times\mathcal{X}\rightarrow[0,\infty)$ is a given distortion measure.
It is well known \cite{CT91} that as the blocklength $n$ approaches infinity, the asymptotic minimum achievable distortion at bit rate $R$  is characterized by 
Shannon's distortion-rate function
\begin{align}
	D(R):=&\inf\limits_{p_{\hat{X}|X}}\mathbb{E}[d(X,\hat{X})]\label{eq:ShannonDR}\\
	&\mbox{s.t.}\quad I(X;\hat{X})\leq R.
\end{align}
However, the distortion level does not always align with human perception of reconstruction quality, especially in the low-resolution regime. To address this discrepancy, it has been proposed to complement the distortion measure 
$d$ with a perception measure 
$\phi$, which typically takes the form of a divergence quantifying the distributional difference between the source and the reconstruction. This further leads to the introduction of the distortion-rate-perception function by Blau and Michaeli \cite{BM19}:
\begin{align}
	D(R,P):=&\inf\limits_{p_{\hat{X}|X}}\mathbb{E}[d(X,\hat{X})]\label{eq:BMDRP}\\
&\mbox{s.t.}\quad I(X;\hat{X})\leq R,\\
&\hspace{0.3in}\phi(p_X,p_{\hat{X}})\leq P.
\end{align}
The coding theorems established by Theis and Wagner \cite{TW21}, as well as by Chen et al. \cite{CYWSGT22}, indicate that $D(R,P)$ retains the operational meaning of $D(R)$ for the lossy source coding system, with the addition of a perception constraint
\begin{align}
	\frac{1}{n}\sum\limits_{t=1}^n\phi(p_{X(t)},p_{\hat{X}(t)})\leq P.\label{eq:symbolP}
\end{align}
The perception constraint above enforces distributional consistency at the symbol level but not at the sequence level. Indeed, setting $P=0$ in \eqref{eq:symbolP} only ensures $p_{\hat{X}(t)}=p_{X(t)}$ for $t=1,2,\ldots,n$, which does not necessarily imply $p_{\hat{X}^n}=p_{X^n}$. To gain more control of the reconstruction distribution, one can replace \eqref{eq:symbolP} with 
\begin{align}
	\frac{1}{n}\phi(p_{X^n},p_{\hat{X}^n})\leq P.\label{eq:sequenceP}
\end{align}
It turns out that this seemingly minor change significantly complicates the problem. From a technical perspective, the new perception constraint is not conducive to single-letterization; in fact, single-letterization appears to be impossible for a generic divergence 
$\phi$. From an operational standpoint, it has been observed  that under the perception constraint \eqref{eq:sequenceP}, the fundamental performance limits are generally not characterized by Blau and Michaeli's distortion-rate-perception function  $D(R,P)$; more surprisingly, they critically depend on the amount of common randomness shared between the encoder and decoder \cite{TA21,LZCK22,LZCK22J,Wagner22}, whereas common randomness is known \cite{CYWSGT22} to have no impact on the performance limits under the perception constraint \eqref{eq:symbolP}. Although some progress has been made by exploring extreme scenarios or alternative formulations \cite{SCKY24, XLCZ24, XLCYZ24}, a comprehensive understanding of the fundamental distortion-rate-perception tradeoff with limited common randomness under the constraint \eqref{eq:sequenceP} is still lacking.

The main contribution of the present work is a complete solution for the special case with the squared error serving as the distortion measure, i.e., $d(x,\hat{x})=\|x-\hat{x}\|^2$, and the squared Wasserstein-2 distance\footnote{It is well known (see, e.g., \cite[Appendix A.2]{FMM21}) that the infimum in \eqref{eq:W2} is attained when $\mathbb{E}[\|X\|^2]<\infty$ and $\mathbb{E}[\|\hat{X}\|^2]<\infty$.} serving as the perception measure, i.e., 
\begin{align}
	\phi(p_X,p_{\hat{X}})=W^2_2(p_X,p_{\hat{X}}):=\inf\limits_{\mu\in\Pi(p_X,p_{\hat{X}})}\mathbb{E}_{\mu}[\|X-\hat{X}\|^2],\label{eq:W2}
\end{align}
where $\|\cdot\|$ denotes the Euclidean norm, and $\Pi(p_X,p_{\hat{X}})$ represents the set of couplings of $p_X$ and $p_{\hat{X}}$.
It will be seen that this solution builds upon the foundational result on the distortion-perception tradeoff in the quadratic Wasserstein space \cite{FMM21}.

The remainder of this paper is organized as follows. Section \ref{sec:review} introduces the problem definition and reviews relevant results from the literature. The main coding theorem is presented in Section \ref{sec:main}, followed by a detailed analysis of the Gaussian case in Section \ref{sec:Gaussian}. Finally, we conclude the paper in Section \ref{sec:conclusion}.

Notation: We use $\mathbb{R}^L$ to represent the $L$-dimensional real vector space and $\mathbb{R}^L_+$ to represent its nonnegative orthant. For any $a,b\in\mathbb{R}$, let $a\vee b:=\max\{a,b\}$, $a\wedge b:=\min\{a,b\}$,  $(a)_+:=\max\{a,0\}$, and 
$\log^+(a):=\max\{\log(a), 0\}$ assuming $a>0$. The set of all probability distributions defined over  $\mathbb{R}^L$ is written as 
$\mathcal{P}(\mathbb{R}^L)$, and the Gaussian distribution with mean $\mu$ and covariance matrix $\Sigma$ is denoted by  
 $\mathcal{N}(\mu,\Sigma)$. For a positive integer $m$,  the set $\{1,2,\ldots,m\}$ is labeled $[m]$, and the uniform distribution over this set is abbreviated as $\mathrm{Unif}[m]$. 
 We use 
$\mathrm{diag}(a_1,a_2,\ldots,a_L)$ to refer to a diagonal matrix with the diagonal entries being $a_1, a_2,\ldots,a_L$. 
Subscripts may be added to information measures or probability operators to specify the underlying distribution when necessary. 
Throughout the paper, $\log(\cdot)$ and $\ln(\cdot)$ denote the logarithms  to base $2$ and base $e$, respectively.

\section{Problem Definition and Literature Review}\label{sec:review}

Let the source $\{X(t)\}_{t=1}^{\infty}$ be an i.i.d. process with marginal distribution $p_X$ over $\mathbb{R}^L$. We assume that $p_X$ is square-integrable, i.e., $\mathbb{E}[\|X\|^2]<\infty$. A length-$n$ perception-aware lossy source coding system consists of a stochastic encoder $f^{(n)}:\mathbb{R}^{L\times n}\times\mathcal{K}\rightarrow\mathcal{J}$, a stochastic decoder $g^{(n)}:\mathcal{J}\times\mathcal{K}\rightarrow\mathbb{R}^{L\times n}$, and a shared random seed $K$, which is uniformly distributed over $\mathcal{K}$ and independent of the source.
The stochastic encoder $f^{(n)}$ maps  the source sequence $X^n$ and random seed $K$ to a codeword $J$ in  $\mathcal{J}$ according to some conditional distribution $p_{J|X^nK}$, while the stochastic decoder $g^{(n)}$ generates $\hat{X}^n$ based on $J$ and $K$ according to some conditional distribution $p_{\hat{X}^n|JK}$. The entire system is charaterized by the joint distribution $p_{X^nJK\hat{X}^n}$, which factors as $p_{X^nJK\hat{X}^n}=p_{X^n}p_Kp_{J|X^nK}p_{\hat{X}^n|JK}$.


\begin{definition}
	A distortion level $D$ is said to be achievable subject to the compression rate constraint $R$, the common randomness rate constraint $C$, and the perception constraint $P$ 	
	if there exists a length-$n$ perception-aware lossy source coding system satisfying	
	\begin{align}
		&\frac{1}{n}\log|\mathcal{J}|\leq R,\label{eq:rate}\\
		&\frac{1}{n}\log|\mathcal{K}|\leq C,\label{eq:commonrandomness}\\
		&\frac{1}{n}\mathbb{E}[\|X^n-\hat{X}^n\|^2]\leq D,\label{eq:distortion}\\
		&\frac{1}{n}W^2_2(p_{X^n},p_{\hat{X}^n})\leq P.\label{eq:perception}
	\end{align}
	The infimum of such achievable $D$ is denoted by  $D^*(R,C,P)$.
\end{definition}

Prior to the present work, a computable characterization of $D^*(R,C,P)$ has only be obtained for certain extreme scenarios, which are reviewed below (see also \cite[Section II-B]{CFKOS25}).

\subsection{No Common Randomness $(C=0)$}\label{subsec:nC}

According to \cite[Theorem 1]{YWL22}, when $C=0$, we can express $D^*(R,C,P)$  in terms of Shannon's distortion-rate function as
\begin{align}
	D^*(R,0,P)=D(R)+[(\sqrt{D(R)}-\sqrt{P})_+]^2.\label{eq:Rc=0}
\end{align}
To gain a better understanding of \eqref{eq:Rc=0}, it is helpful to consider the following problem. Let $S$ be an $N$-dimensional random vector satisfying $\mathbb{E}[\|S\|^2]<\infty$, and let $W$ be a representation of $S$ generated through some conditional distribution $p_{W|S}$. The goal is to produce a reconstruction $\hat{S}$ based on $W$ that optimally balances the tradeoff between distortion loss $\mathbb{E}[\|S-\hat{S}\|^2]$
and perception loss $W^2_2(p_S,p_{\hat{S}})$. As shown in \cite[Theorem 1]{FMM21}, if $W^2_2(p_S,p_{\hat{S}})\leq P$, then
\begin{align}
	\mathbb{E}[\|S-\hat{S}\|^2]\geq\mathbb{E}[\|S-\tilde{S}\|^2]+[(W_2(p_S,p_{\tilde{S}})-\sqrt{P})_+]^2,\label{eq:DPlowerbound}
\end{align}
where $\tilde{S}:=\mathbb{E}[S|W]$. Moreover, the lower bound in \eqref{eq:DPlowerbound} is achieved by an interpolation scheme \cite[Theorem 3]{FMM21}. Specifically, let\footnote{Here, we assume $W_2(p_S,p_{\tilde{S}})>0$; otherwise, $\tilde{S}=S$ a.s.} 
\begin{align}
	\hat{S}^*:=\begin{cases}
		\left(1-\frac{\sqrt{P}}{W_2(p_S,p_{\tilde{S}})}\right)S'+\frac{\sqrt{P}}{W_2(p_S,p_{\tilde{S}})}\tilde{S}, & P\in[0,W^2_2(p_S,p_{\tilde{S}})],\\
		\tilde{S}, & P>W^2_2(p_S,p_{\tilde{S}}),
	\end{cases}
\end{align}
where $S'$ is jointly distributed with $(S,W)$ such that $S\leftrightarrow W\leftrightarrow S'$ forms a Markov chain, $p_{S'}=p_S$, and $\mathbb{E}[\|S'-\tilde{S}\|^2]=W^2_2(p_S,p_{\tilde{S}})$. In other words, $\hat{S}^*$ is a linear combination of the MMSE estimate $\tilde{S}$ and a perceptually perfect reconstruction $S'$ generated from $\tilde{S}$ 
through optimal transport. It achieves the best possible distortion-perception tradeoff for the given representation $W$
since
\begin{align}
	W^2_2(p_S,p_{\hat{S}^*})=W^2_2(p_S,p_{\tilde{S}})\wedge P\leq P\label{eq:S*1}
\end{align}
and
\begin{align}
	\mathbb{E}[\|S-\hat{S}^*\|^2]=\mathbb{E}[\|S-\tilde{S}\|^2]+[(W_2(p_S,p_{\tilde{S}})-\sqrt{P})_+]^2.\label{eq:S*2}
\end{align}

Note that in the context of perception-aware lossy source coding, $S$, $W$, $\tilde{S}$, and $\hat{S}$ correspond to $X^n$, $(J,K)$, $\mathbb{E}[X^n|J,K]$, and $\hat{X}^n$, respectively. Therefore, given the encoder and the perception constraint \eqref{eq:perception}, the optimal decoder is fully specified, and the resulting end-to-end distortion is given by
\begin{align}
\frac{1}{n}\mathbb{E}[\|X^n-\hat{X}^n\|^2]=\frac{1}{n}\mathbb{E}[\|X^n-\tilde{X}^n\|^2]+\left[\left(\frac{1}{\sqrt{n}}W_2(p_{X^n},p_{\tilde{X}^n})-\sqrt{P}\right)_+\right]^2,\label{eq:nDRP}
\end{align}
where $\tilde{X}^n:=\mathbb{E}[X^n|J,K]$. The remaining task is to optimize the encoder. When $C=0$, the random seed $K$ is a constant and can be ignored. Let $D^{(n)}(R)$ denote the minimum achievable distortion $\frac{1}{n}\mathbb{E}[\|X^n-\tilde{X}^n\|^2]$ under the rate constraint \eqref{eq:rate}. The key observation is that, in the absence of common randomness, 
the minimum achievable $\frac{1}{n}W^2_2(p_{X^n},p_{\tilde{X}^n})$ under the same rate constraint  is also $D^{(n)}(R)$ and is  attained by the same $\tilde{X}^n$ that achieves $D^{(n)}(R)$. This is because if $\frac{1}{n}W^2_2(p_{X^n},p_{\tilde{X}^n})<D^{(n)}(R)$, then the associated optimal transport plan would imply the existence of a lossy source coding system with rate $R$ and  end-to-end distortion strictly lower than $D^{(n)}(R)$, leading to a contradiction. As a consequence, substituting $\frac{1}{n}\mathbb{E}[\|X^n-\tilde{X}^n\|^2]$ and $\frac{1}{n}W^2_2(p_{X^n},p_{\tilde{X}^n})$
with $D^{(n)}(R)$ in \eqref{eq:nDRP} yields the minimum achievable distortion level for a length-$n$ perception-aware lossy source coding system satisfying \eqref{eq:rate}, \eqref{eq:commonrandomness} with $C=0$, and \eqref{eq:perception}:
\begin{align}
	D^{(n)}(R)+[(\sqrt{D^{(n)}(R)}-\sqrt{P})_+]^2.
\end{align}
Finally,  invoking the fact that $D^{(n)}(R)\rightarrow D(R)$ as $n\rightarrow\infty$ proves \eqref{eq:Rc=0}.

One remarkable feature of the above coding system is that the encoder only needs to minimize distortion and can be designed without knowledge of the perception constraint. In this sense, the output of an optimal conventional lossy source encoder  provides a universally applicable representation in the perception-aware setting. We refer to such a representation as {\em one-shot universal} since it is associated with a fixed blocklength. Notably, this universality is established under the assumption that no common randomness is available. 
In contrast, the decoder must adjust the interpolation coefficients according to the perception constraint.  As
\begin{align} \frac{1}{n}\mathbb{E}[\|X^n-\tilde{X}^n\|^2]=\frac{1}{n}W^2_2(p_{X^n},p_{\tilde{X}^n})=D^{(n)}(R),
\end{align}	
the optimal transport plan for generating 
a perceptually perfect reconstruction based on the MMSE estimate $\tilde{X}^n$ can be realized through posterior sampling \cite{YWYML21}.

\subsection{Unlimited Common Randomness $(C=\infty)$}

It is shown in \cite[Theorem 7]{CYWSGT22} that 
\begin{align}
	D^*(R,\infty,P)=D(R,P),\label{eq:Rcinfty}
\end{align}
i.e., $D^*(R,C,P)$ coincides with Blau and Michaeli's distortion-rate-perception function when $C=\infty$. An intuitive explanation is as follows. With unlimited common randomness, there is no loss of optimality in restricting the reconstruction sequence $\hat{X}^n$ to be i.i.d. Consequently, by the tensorization property of the squared Wasserstein-2 distance, we have
\begin{align}
	\frac{1}{n}W^2_2(p_{X^n},p_{\hat{X}^n})=\frac{1}{n}\sum\limits_{t=1}^nW^2_2(p_{X(t)},p_{\hat{X}(t)}).
\end{align}
The perception constraint in \eqref{eq:perception} then reduces to
\begin{align}
	\frac{1}{n}\sum\limits_{t=1}^nW^2_2(p_{X(t)},p_{\hat{X}(t)})\leq P,
\end{align}
which takes the same form as \eqref{eq:symbolP}. Hence, the fundamental distortion-rate-perception tradeoff is characterized by $D(R,P)$.

\subsection{Perfect Perception $(P=0)$}

Note that $P=0$ forces $p_{\hat{X}^n}=p_{X^n}$, meaning the reconstruction sequence must be i.i.d. with the same marginal distribution as the source. This makes perception-aware lossy source coding a special case of output-constrained lossy source coding. As a result, the coding theorem  \cite[Theorem 2]{Wagner22} \cite[Theorem 1]{SLY15J2} for the latter can be readily applied, yielding
\begin{align}
	D^*(R,C,0)=&\inf\limits_{p_{U\hat{X}|X}}\mathbb{E}[\|X-\hat{X}\|^2]\label{eq:saldi1}\\
	&\mbox{s.t.}\quad I(X;U)\leq R,\label{eq:saldi2}\\
	&\hspace{0.3in}I(\hat{X};U)\leq R+C,\label{eq:saldi3}\\
	&\hspace{0.3in}X\leftrightarrow U\leftrightarrow\hat{X} \mbox{ forms a Markov chain},\label{eq:saldi4}\\
	&\hspace{0.3in}p_{\hat{X}}=p_X.\label{eq:saldi5}
\end{align}

It can be verified that the characterizations of 
$D^*(R,C,P)$ in the three aforementioned extreme scenarios are consistent. However, they take very different forms, suggesting that unification is not straightforward.

\section{Coding Theorem}\label{sec:main}

The main result of this section is a single-letter characterization of $D^*(R,C,P)$. To this end, we introduce 
\begin{align}
	D(R,C,P):=&\inf\limits_{p_{\tilde{X}}\in\mathcal{P}(\mathbb{R}^L), \mu,\nu\in\Pi(p_X,p_{\tilde{X}})}\mathbb{E}_{\mu}[\|X-\tilde{X}\|^2]+\left[\left(\sqrt{\mathbb{E}_{\nu}[\|X-\tilde{X}\|^2]}-\sqrt{P}\right)_+\right]^2\label{eq:inf}\\
	&\mbox{s.t.}\quad\mathbb{E}_{\mu}[X|\tilde{X}]=\tilde{X}\quad\mu-\mbox{a.s.},\label{eq:conditionexp}\\
	&\hspace{0.3in}I_{\mu}(X;\tilde{X})\leq R,\label{eq:constraintR}\\
	&\hspace{0.3in}I_{\nu}(X;\tilde{X})\leq R+C.\label{eq:constraintRc}	
\end{align}
As shown below, the infimum in \eqref{eq:inf} is in fact a minimum.

\begin{proposition}\label{prop:inf}
	For $p_X$ with $\mathbb{E}[\|X\|^2]<\infty$,  the infimum in \eqref{eq:inf} is attained.
\end{proposition}
\begin{IEEEproof}
	See Appendix \ref{app:inf}.
	\end{IEEEproof}

The following proposition delineates several fundamental properties of $D(R,C,P)$.

\begin{proposition}\label{prop:convexity}
For $p_X$ with $\mathbb{E}[\|X\|^2]<\infty$, the function $D(R,C,P)$ is decreasing, convex, and continuous in $(R,C,P)$.
\end{proposition}
\begin{IEEEproof}
	See Appendix \ref{app:convexity}.
	\end{IEEEproof}



Having established the required preliminaries, we now turn to our main result.
\begin{theorem}\label{thm:main}
	For $p_X$ with $\mathbb{E}[\|X\|^2]<\infty$,
	\begin{align}
		D^*(R,C,P)=D(R,C,P).
	\end{align}
\end{theorem}
\begin{IEEEproof}
Here we just give a sketch of the proof. The details are relegated to Appendix \ref{app:main}. Given $p_{\tilde{X}}\in\mathcal{P}(\mathbb{R}^L)$ and $\mu,\nu\in\Pi(p_X,p_{\tilde{X}})$ satisfying
\eqref{eq:conditionexp}--\eqref{eq:constraintRc}, we show, via a soft covering lemma with respect to the Wasserstein-2 distance, that there exists an encoder satisfying
\begin{align}
	&\frac{1}{n}\log|\mathcal{J}|\approx R,\\
	&\frac{1}{n}\log|\mathcal{K}|\approx C,\\
	&\frac{1}{n}\mathbb{E}[\|X^n-\tilde{X}^n\|^2]\lessapprox\mathbb{E}_{\mu}[\|X-\tilde{X}\|^2],\\
	&\frac{1}{n}W^2_2(p_{X^n},\tilde{X}^n)\lessapprox\mathbb{E}_{\nu}[\|X-\tilde{X}\|^2],
\end{align}
where $\tilde{X}^n:=\mathbb{E}[X^n|J,K]$. Employing the optimal decoder based on the linear interpolation scheme described in Section \ref{subsec:nC} yields 
\begin{align}
\frac{1}{n}\mathbb{E}[\|X^n-\hat{X}^n\|^2]&=\frac{1}{n}\mathbb{E}[\|X^n-\tilde{X}^n\|^2]+\left[\left(\frac{1}{\sqrt{n}}W_2(p_{X^n},p_{\tilde{X}^n})-\sqrt{P}\right)_+\right]^2\nonumber\\
&\lessapprox\mathbb{E}_{\mu}[\|X-\tilde{X}\|^2]+\left[\left(\sqrt{\mathbb{E}_{\nu}[\|X-\tilde{X}\|^2]}-\sqrt{P}\right)_+\right]^2.
\end{align}
We then complete the proof by establishing a matching converse.
\end{IEEEproof}

Now we proceed to show that Theorem \ref{thm:main} recovers existing results in the extreme scenarios.
\begin{enumerate}
	\item When $C=0$, the constraint \eqref{eq:conditionexp} becomes redundant. In this case,
both $\mathbb{E}_{\mu}[\|X-\tilde{X}\|^2]$ and $\mathbb{E}_{\nu}[\|X-\tilde{X}\|^2]$ attain the same minimum value $D(R)$, leading to the expression in \eqref{eq:Rc=0}. 

\item When $C=\infty$, \eqref{eq:inf}--\eqref{eq:constraintRc} simplifies to
\begin{align}
	D(R,\infty,P)=&\inf\limits_{p_{\hat{X}}\in\mathcal{P}(\mathbb{R}^L), \mu\in\Pi(p_X,p_{\tilde{X}})}\mathbb{E}_{\mu}[\|X-\tilde{X}\|^2]+[(W_2(p_X,p_{\tilde{X}})-\sqrt{P})_+]^2\\
&\mbox{s.t.}\quad\mathbb{E}_{\mu}[X|\tilde{X}]=\tilde{X}\quad\mu-\mbox{a.s.},\label{eq:inftymmse}\\
&\hspace{0.3in}I_{\mu}(X;\tilde{X})\leq R,\label{eq:inftyR}	
\end{align}
which is equivalent to \eqref{eq:Rcinfty} for the following reason. Given $p_{\hat{X}|X}$ satisfying $I(X;\hat{X})\leq R$ and $W^2_2(p_X,p_{\hat{X}})\leq P$, it follows by the data processing inequality and \eqref{eq:DPlowerbound} that
\begin{align}
	&I(X;\tilde{X})\leq R,\\
	&\mathbb{E}[\|X-\hat{X}\|^2]\geq\mathbb{E}[\|X-\tilde{X}\|^2]+[(W_2(p_X,p_{\tilde{X}})-\sqrt{P})_+]^2,
\end{align}
where $\tilde{X}:=\mathbb{E}[X|\hat{X}]$. This implies $D(R,P)\geq D(R,\infty,P)$. Conversely, given $p_{\tilde{X}|X}$ satisfing \eqref{eq:inftymmse} and \eqref{eq:inftyR}, let
\begin{align}
	\hat{X}:=\begin{cases}
		\left(1-\frac{\sqrt{P}}{W_2(p_X,p_{\tilde{X}})}\right)X'+\frac{\sqrt{P}}{W_2(p_X,p_{\tilde{X}})}\tilde{X}, & P\in[0,W^2_2(p_X,p_{\tilde{X}})],\\
		\tilde{X}, & P>W^2_2(p_X,p_{\tilde{X}}),
	\end{cases}
\end{align}
where $X'$ is jointly distributed with $(X,\tilde{X})$ such that $X\leftrightarrow \tilde{X}\leftrightarrow X'$ forms a Markov chain, $p_{X'}=p_X$, and $\mathbb{E}[\|X'-\tilde{X}\|^2]=W^2_2(p_X,p_{\tilde{X}})$; in light of the data processing inequality, \eqref{eq:S*1}, and \eqref{eq:S*2},
\begin{align}
	&I(X;\hat{X})\leq R,\\
	&W^2_2(p_X,p_{\hat{X}})\leq P,\\
	&\mathbb{E}[\|X-\hat{X}\|^2]=\mathbb{E}[\|X-\tilde{X}\|^2]+[(W_2(p_X,p_{\tilde{X}})-\sqrt{P})_+]^2.
\end{align}
Therefore, we  also have $D(R,P)\leq D(R,\infty,P)$. 

\item When $P=0$,  \eqref{eq:inf}--\eqref{eq:constraintRc} becomes
\begin{align}
	D(R,C,0)=&\inf\limits_{p_{\tilde{X}}\in\mathcal{P}(\mathbb{R}^L), \mu,\nu\in\Pi(p_X,p_{\tilde{X}})}\mathbb{E}_{\mu}[\|X-\tilde{X}\|^2]+\mathbb{E}_{\nu}[\|X-\tilde{X}\|^2]\\
	&\mbox{s.t.}\quad\mathbb{E}_{\mu}[X|\tilde{X}]=\tilde{X}\quad\mu-\mbox{a.s.},\label{eq:becomes1}\\
	&\hspace{0.3in}I_{\mu}(X;\tilde{X})\leq R,\label{eq:becomes2}\\
	&\hspace{0.3in}I_{\nu}(X;\tilde{X})\leq R+C.\label{eq:becomes3}
\end{align}
Given $p_{U\hat{X}|X}$ satisfying \eqref{eq:saldi2}--\eqref{eq:saldi5},
let $\tilde{X}:=\mathbb{E}[X|U]$. By the data processing inequality,
\begin{align}
	&I(X;\tilde{X})\leq R,\\
	&I(\hat{X};\tilde{X})\leq R+C.
\end{align}
Moreover, we have
\begin{align}
	\mathbb{E}[\|X-\hat{X}\|^2]=\mathbb{E}[\|X-\tilde{X}\|^2]+\mathbb{E}[\|\hat{X}-\tilde{X}\|^2].
\end{align}
Since both $p_{X\tilde{X}}$ and $p_{\hat{X}\tilde{X}}$ belong to $\Pi(p_X,p_{\hat{X}})$, it follows that $D(R,C,0)$ is
upper bounded by the expression in \eqref{eq:saldi1}.
Conversely, given $\mu,\nu\in\Pi(p_X,p_{\tilde{X}})$ satisfying \eqref{eq:becomes1}--\eqref{eq:becomes3}, construct $p_{XU\hat{X}}$ such that $X\leftrightarrow U\leftrightarrow\hat{X}$ forms a Markov chain, with $p_{XU}=\mu$ and $p_{\hat{X}U}=\nu$. Under this construction, the conditions  \eqref{eq:saldi2}--\eqref{eq:saldi5} hold, and in view of  \eqref{eq:becomes1},
\begin{align}
	\mathbb{E}[\|X-\hat{X}\|^2]&=\mathbb{E}[\|X-U\|^2]+\mathbb{E}[\|\hat{X}-U\|^2]\nonumber\\
	&=\mathbb{E}_{\mu}[\|X-\tilde{X}\|^2]+\mathbb{E}_{\nu}[\|X-\tilde{X}\|^2].
\end{align}
As a consequence, $D(R,C,0)$ is also
lower bounded by the expression in \eqref{eq:saldi1}.
\end{enumerate}




Note that if the minimizer of the optimization problem associated with 
$D(R,C,P)$ does not depend on $P$, then, operationally, as the blocklength 
$n\rightarrow\infty$, the encoder can be designed without knowledge of $P$, which implies the existence of coded representations that are {\em asymptotically universal}. This is indeed the case when 
$C=0$. In fact, as discussed in Section \ref{subsec:nC}, without common randomness, universal representations can even be found in the one-shot setting. However, when 
$C>0$, it is not immediately clear whether such representations, even in the asymptotic sense, exist. With this question in mind, we turn to the next section.


\section{Gaussian Case}\label{sec:Gaussian}

This section focuses on the case where $p_X$ is a Gaussian distribution, allowing for an explicit solution to the optimization problem associated with  $D(R,C,P)$. As we will see, the solution sheds light on the existence of universal representations in the presence of limited common randomness.

\subsection{Scalar Source}

The following result provides an explicit characterization of $D^*(R,C,P)$ for the scalar Gaussian source, unifying and generalizing Zhang et al.’s result for the unlimited common randomness scenario (i.e., $C=\infty$) \cite{ZQCK21} and Wagner’s result for the perfect perception scenario (i.e., $P=0$) \cite{Wagner22}. Let
\begin{align}
	\psi(a,b):=\sqrt{(1-2^{-2a})(1-2^{-2b})}.
\end{align}
\begin{theorem}\label{thm:Gaussian}
	For $p_X=\mathcal{N}(0,\gamma)$, 
	\begin{align}
		D^*(R,C,P)=\gamma 2^{-2R}+\left[\left(\sqrt{\gamma(2-2^{-2R}-2\psi(R,R+C))}-\sqrt{P}\right)_+\right]^2.
	\end{align}
\end{theorem}
\begin{IEEEproof}
	First we shall show that
	\begin{align}
		D(R,C,P)\leq\gamma 2^{-2R}+\left[\left(\sqrt{\gamma(2-2^{-2R}-2\psi(R,R+C))}-\sqrt{P}\right)_+\right]^2.\label{eq:prove1}
	\end{align}
Let $p_{\tilde{X}}$ be a Gaussian distribution with mean zero and variance $\gamma(1-2^{-2R})$. Moreover, let $\mu$ and $\nu$ be zero-mean bivariate Gaussian distributions with covariance matrices
\begin{align}
	\left(\begin{matrix}
		\gamma & \gamma(1-2^{-2R})\\
		\gamma(1-2^{-2R}) & \gamma(1-2^{-2R})
	\end{matrix}\right)
\end{align}
and
\begin{align}
	\left(\begin{matrix}
		\gamma & \gamma\psi(R,R+C)\\
		\gamma\psi(R,R+C) & \gamma(1-2^{-2R})
	\end{matrix}\right),
\end{align}
respectively. Clearly, $\mu,\nu\in\Pi(p_X,p_{\tilde{X}})$. In addition, it can be verified that
\begin{align}
	&\mathbb{E}_{\mu}[X|\tilde{X}]=\tilde{X}\quad\mu-\mbox{a.s.},\\
	&I_{\mu}(X;\tilde{X})=R,\\
	&I_{\nu}(X;\tilde{X})=R+C,\\
	&\mathbb{E}_{\mu}[(X-\tilde{X})^2]=\gamma 2^{-2R},\\
	&\mathbb{E}_{\nu}[(X-\tilde{X})^2]=\gamma(2-2^{-2R}-2\psi(R,R+C)).
\end{align}
This proves \eqref{eq:prove1}.

Next we proceed to show that
\begin{align}
	D(R,C,P)\geq\gamma 2^{-2R}+\left[\left(\sqrt{\gamma(2-2^{-2R}-2\psi(R,R+C))}-\sqrt{P}\right)_+\right]^2.\label{eq:prove2}
\end{align}
Consider $p_{\tilde{X}}\in\mathcal{P}(\mathbb{R})$ and $\mu,\nu\in\Pi(p_X,p_{\tilde{X}})$  satisfying
\begin{align}
		&\mathbb{E}_{\mu}[X|\tilde{X}]=\tilde{X}\quad\mu-\mbox{a.s.},\label{eq:as}\\
	&I_{\mu}(X;\tilde{X})\leq R,\label{eq:Rconstraint}\\
	&I_{\nu}(X;\tilde{X})\leq R+C.\label{eq:R+Cconstraint}
\end{align}
Let $\tilde{\gamma}:=\mathbb{E}[\tilde{X}^2]$ and $\theta:=\mathbb{E}_{\nu}[X\tilde{X}]$. It follows by \eqref{eq:as}  that 
\begin{align}
	\mathbb{E}_{\mu}[(X-\tilde{X})^2]=\gamma-\tilde{\gamma},\label{eq:dmu}
\end{align}
and consequently,
\begin{align}
	h_{\mu}(X-\tilde{X})\leq\frac{1}{2}\log(2\pi e(\gamma-\tilde{\gamma})).
\end{align}
Therefore, we have
\begin{align}
I_{\mu}(X;\tilde{X})&=h(X)-h_{\mu}(X|\tilde{X})\nonumber\\
&\geq h(X)-h_{\mu}(X-\tilde{X})\nonumber\\
	&\geq\frac{1}{2}\log\left(\frac{\gamma}{\gamma-\tilde{\gamma}}\right),
\end{align}
which, together with \eqref{eq:Rconstraint}, implies
\begin{align}
	\tilde{\gamma}\leq\gamma (1-2^{-2R}).\label{eq:tildegamma}
\end{align}
Substituting \eqref{eq:tildegamma} into \eqref{eq:dmu} gives
\begin{align}
	\mathbb{E}_{\mu}[(X-\tilde{X})^2]\geq\gamma 2^{-2R}.\label{eq:term1}
\end{align}
Moreover, we have
\begin{align}
I_{\nu}(X;\tilde{X})&=h(X)-h_{\nu}(X|\tilde{X})\nonumber\\
&\geq h(X)-h_{\nu}\left(X-\frac{\theta}{\tilde{\gamma}}\tilde{X}\right)\nonumber\\
&\geq\frac{1}{2}\log\left(\frac{\gamma\tilde{\gamma}}{\gamma\tilde{\gamma}-\theta^2}\right),
\end{align}
which, together with \eqref{eq:R+Cconstraint}, implies
\begin{align}
	\theta\leq\sqrt{\gamma\tilde{\gamma}(1-2^{-2(R+C)})}.
\end{align}
As a consequence, 
\begin{align}
	\mathbb{E}_{\nu}[(X-\tilde{X})^2]&=\gamma+\tilde{\gamma}-2\theta\nonumber\\
	&\geq\gamma+\tilde{\gamma}-2\sqrt{\gamma\tilde{\gamma}(1-2^{-2(R+C)})}\nonumber\\
	&\geq\gamma(2-2^{-2R}-2\psi(R,R+C)),\label{eq:term2}
\end{align}
where the last inequality is due to \eqref{eq:tildegamma} and the fact that $\gamma+\tilde{\gamma}-2\sqrt{\gamma\tilde{\gamma}(1-2^{-2(R+C)})}$ is monotonically decreasing in $\tilde{\gamma}$ for $\tilde{\gamma}\in[0,\gamma(1-2^{-2(R+C)})]$. Combining \eqref{eq:term1} and \eqref{eq:term2} proves \eqref{eq:prove2}.

In view of  \eqref{eq:prove1} and \eqref{eq:prove2}, the proof is complete by invoking Theorem \ref{thm:main}.
	\end{IEEEproof}

As shown in the proof of Theorem \ref{thm:Gaussian},  for the scalar Gaussian source, the minimizer of the optimization problem associated with $D(R,C,P)$ does not depend on $P$. Consequently,  there exist coded representations that are asymptotically universal in the large blocklength limit. This generalizes the finding in \cite{ZQCK21} for the unlimited common randomness scenario (i.e., $C=\infty$).

\begin{figure}[htbp]
	\centerline{\includegraphics[width=10cm]{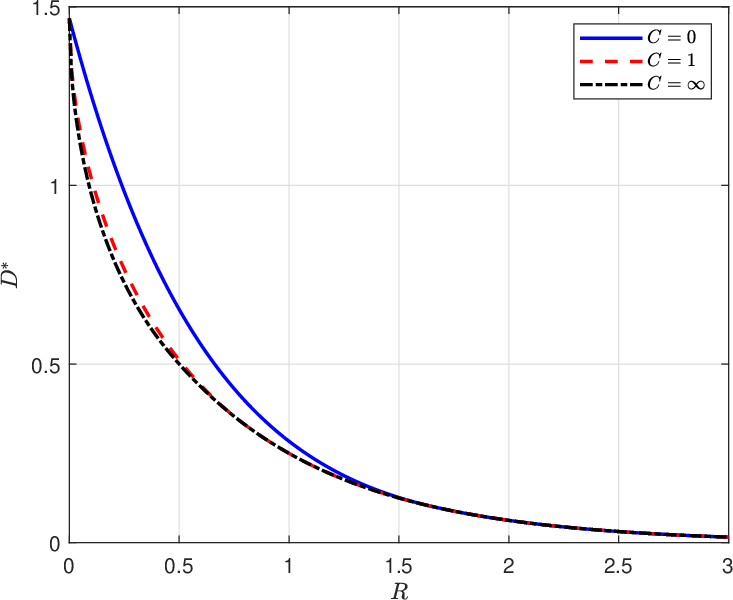}} \caption{Illustration of $D^*(R,C,P)$ as a function of $R$, for $P=0.1$ and different values of $C$, with $p_X=\mathcal{N}(0,1)$.}
	\label{fig:rdp1} 
\end{figure}

\begin{figure}[htbp]
	\centerline{\includegraphics[width=10cm]{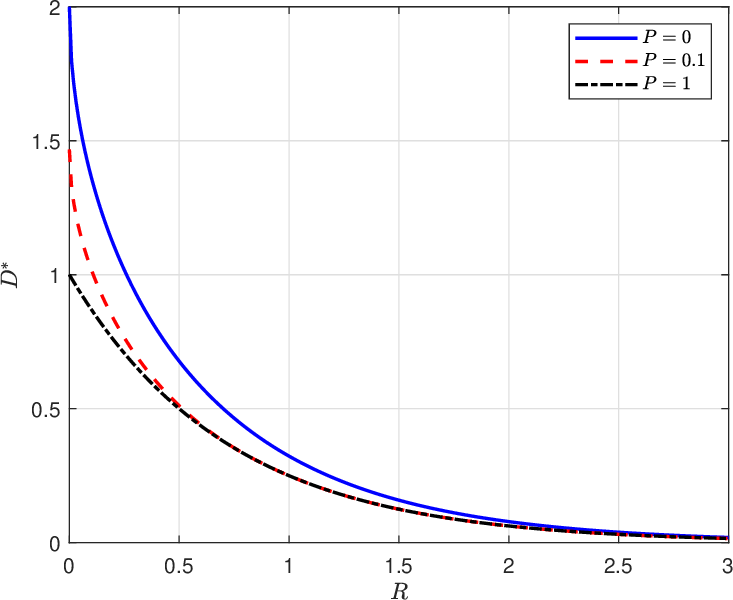}} \caption{Illustration of $D^*(R,C,P)$ as a function of $R$, for $C=1$ and different values of $P$, with $p_X=\mathcal{N}(0,1)$.}
	\label{fig:rdp2} 
\end{figure}

In view of Theorem \ref{thm:Gaussian}, when
\begin{align}
	P\geq\gamma\left(2-2^{-2R}-2\sqrt{(1-2^{-2R})(1-2^{-2(R+C)})}\right),\label{eq:Pthreshold}
\end{align}
the perception constraint is inactive, and 
\begin{align}
	D^*(R,C,P)=\gamma 2^{-2R},
\end{align}
degenerating to the classical quadratic Gaussian distortion-rate function. Note that \eqref{eq:Pthreshold} can be written equivalently as
\begin{align}
C\geq\begin{cases}
	\frac{1}{2}\log\left(\frac{4(2^{-2R}-2^{-4R})}{2(2-2^{-2R})\frac{P}{\gamma}-\frac{P^2}{\gamma^2}-2^{-4R}}\right), & P\in(\gamma(2-2^{-2R}-2\sqrt{1-2^{-2R}}),\gamma 2^{-2R}),\\
	0, & P\geq\gamma 2^{-2R}\mbox{ or }R=0,	
\end{cases}
\end{align}
which highlights the regime where increasing $C$ alone does not further reduce $D^*(R,C,P)$. It is also possible to express \eqref{eq:Pthreshold} as 
 a threshold on $R$:
\begin{align}
	R\geq\begin{cases}
		\frac{1}{2}\log\left(\frac{\frac{2P}{\gamma}+2^{-2(C-1)}+\sqrt{(\frac{2P}{\gamma}+2^{-2(C-1)})^2+4(1-2^{-2(C-1)})(\frac{4P}{\gamma}-\frac{P^2}{\gamma^2})}}{2(\frac{4P}{\gamma}-\frac{P^2}{\gamma^2})}\right),& P\in(0,\gamma),\\
		0,& P\geq\gamma.
	\end{cases}\label{eq:Rthreshold}
\end{align}

To gain a more concrete understanding, we present some numerical plots of $D^*(R,C,P)$ for $p_X=\mathcal{N}(0,1)$.  Fig. \ref{fig:rdp1} illustrates three distortion-rate curves: the blue curve corresponds to $C=0$ (no common randomness), the red curve  to $C=1$ (limited common randomness), and the black curve to $C=\infty$ (unlimited common randomness).
Setting $\gamma=1$, $C=\infty$, and $P=0.1$ in \eqref{eq:Rthreshold}, we find that the black curve reduces to the classical distortion-rate function for rates above the threshold $R\approx 0.45$. In comparison, the thresholds for the red and blue curves are $R\approx 0.81$ and $R\approx 1.66$, respectively. 
Thus, the red and black curves coincide for $R\gtrapprox 0.81$,  and all three curves coincide for $R\gtrapprox 1.66$. As evident from Fig. \ref{fig:rdp1}, the red curve generally tracks the black curve closely, suggesting that even a small amount of common randomness can be nearly as effective as unlimited common randomness. In Fig. \ref{fig:rdp2}, we fix $C=1$ and plot the distortion-rate curves for three values of $P$: $P=0$ (blue), $P=0.1$ (red), and $P=1$ (black). These different values of $P$ result in distinct starting points for the curves. Specifically, since $D^*(0,C,P)=\gamma+(\sqrt{\gamma}-\sqrt{P})^2=2-2\sqrt{P}+P$, we have $D^*(0,C,0)=2$,  $D^*(0,C,0.1)\approx1.47$, and $D^*(0,C,1)=1$.
When $P=1$, the perception constraint is inactive, so the black curve exactly follows the classical quadratic Gaussian distortion-rate function. In contrast, the perception constraint is always active when $P=0$. When $P=0.1$, the perception constraint becomes inactive for 
$R\gtrapprox0.81$, where the red curve coincides with the black one.

\subsection{Vector Source}

Now consider the vector Gaussian source setting, where $X:=(X_1,X_2,\ldots,X_L)^T\sim\mathcal{N}(0,\Gamma)$. Owing to the invariance of the Euclidean distance under unitary transformations, we can, without loss of generality, diagonalize the covariance matrix via eigenvalue decomposition and assume $\Gamma=\mathrm{diag}(\gamma_1,\gamma_2,\ldots,\gamma_L)$. Under this assumption,  $X_1,X_2,\ldots,X_L$ are mutually independent. The following result generalizes \cite[Theorem 5]{QSCKYSGT25}, which is restricted to the unlimited common randomness scenario (i.e., $C=\infty$).
\begin{theorem}\label{thm:Gaussianvector}
	For $p_X=\mathcal{N}(0,\Gamma)$,
	\begin{align}
		D^*(R,C,P)=&\min\limits_{(r_1,r_2,\ldots,r_L),(r'_1,r'_2,\ldots,r'_L)\in\mathbb{R}^L_+}\sum\limits_{\ell=1}^L\gamma_{\ell} 2^{-2r_{\ell}}+\left[\left(\sqrt{\sum\limits_{\ell=1}^L\gamma_{\ell}(2-2^{-2r_{\ell}}-2\psi(r_{\ell},r'_{\ell}))}-\sqrt{P}\right)_+\right]^2\\
		&\mbox{s.t.}\quad \sum\limits_{\ell=1}^Lr_{\ell}\leq R,\label{eq:sum1}\\
		&\hspace{0.3in}\sum\limits_{\ell=1}^Lr'_{\ell}\leq R+C.\label{eq:sum2}
	\end{align}
\end{theorem}
\begin{IEEEproof}
The result follows by a straightforward extension of the proof for the scalar case; detailed arguments are provided in Appendix \ref{app:Gaussianvector}.
	\end{IEEEproof}

It is natural to ask whether the existence of universal representations extends to the vector Gaussian source setting. 
Specializing Theorem \ref{thm:Gaussianvector} to the case of $P=\infty$ gives
\begin{align}
	D^*(R,C,\infty)=&\min\limits_{(r_1,r_2,\ldots,r_{L}),(r'_1,r'_2,\ldots,r'_{L})\in\mathbb{R}_+^L}\sum\limits_{\ell=1}^L\gamma_{\ell}2^{-2r_{\ell}}\label{eq:min}\\
	&\mbox{s.t.}\quad\sum\limits_{\ell=1}^Lr_{\ell}\leq R,\\
	&\hspace{0.3in}\sum\limits_{\ell=1}^Lr'_{\ell}\leq R+C.
\end{align}
The minimizer of the above optimization problem is given by the classical reverse waterfilling formula
\begin{align}
	r_{\ell}=\hat{r}_{\ell}:=\frac{1}{2}\log^+\left(\frac{\gamma_{\ell}}{\alpha}\right),\quad\ell=1,2,\ldots,L,
\end{align}
with $\alpha$ being the unique number in $(0,\max\{\gamma_{1},\gamma_2,\ldots,\gamma_L\}]$ satisfying
\begin{align}
\sum\limits_{\ell=1}^L\frac{1}{2}\log^+\left(\frac{\gamma_{\ell}}{\alpha}\right)=R,
\end{align}
while the vector $(r'_1,r'_2,\ldots,r'_{L})$ can be chosen arbitrarily  in $\mathbb{R}^L_+$ subject to the constraint $\sum_{\ell=1}^Lr'_{\ell}\leq R+C$.
Note that $(\hat{r}_1,\hat{r}_2,\ldots,\hat{r}_L)$ does not depend on the value of $C$.
On the other hand,  
\begin{align}
	D^*(R,C,0)=&\min\limits_{(r_1,r_2,\ldots,r_L),(r'_1,r'_2,\ldots,r'_L)\in\mathbb{R}^L_+}\sum\limits_{\ell=1}^L2\gamma_{\ell}\left(1-\sqrt{(1-2^{-2r_{\ell}})(1-2^{-2r'_{\ell}})}\right)\label{eq:minimum}\\
	&\mbox{s.t.}\quad\sum\limits_{\ell=1}^Lr_{\ell}\leq R,\\
	&\hspace{0.3in}\sum\limits_{\ell=1}^Lr'_{\ell}\leq R+C.
\end{align}
In general, the minimizer of \eqref{eq:minimum} is different from that of \eqref{eq:min} when $C>0$. This can be seen easily by considering the unlimited common randomness scenario (i.e., $C=\infty$).
When $C=\infty$, there is no loss of optimality in sending $r'_{\ell}$ to infinity, $\ell=1,2,\ldots,L$, and the minimum in \eqref{eq:minimum} is attained  at 
\begin{align}
	r_{\ell}=\check{r}_{\ell}:=\frac{1}{2}\log\left(\frac{1+\sqrt{1+\beta\gamma^2_{\ell}}}{2}\right),
\end{align}
with $\beta$ being the unique nonnegative number satisfying
\begin{align}
	\sum\limits_{\ell=1}^L\frac{1}{2}\log\left(\frac{1+\sqrt{1+\beta\gamma^2_{\ell}}}{2}\right)=R.
\end{align}
It is clear that $(\hat{r}_1,\hat{r}_2,\ldots,\hat{r}_L)$ and $(\check{r}_1,\check{r}_2,\ldots,\check{r}_L)$ are  generally different. Hence, for the vector Gaussian source, universal representations, even in the asymptotic sense, do not necessarily exist when $C>0$.










\section{Conclusion}\label{sec:conclusion}

We have derived a single-letter characterization of the fundamental distortion-rate-perception tradeoff  under the squared error distortion measure and the squared Wasserstein-2 perception measure, in the presence of limited common randomness. Extending this characterization to other distortion and perception measures represents a natural direction for future research. Notably, our analysis relies heavily on properties unique to the quadratic Wasserstein space, suggesting that new analytical tools will be required for broader generalizations.

Beyond its theoretical contributions, our work offers valuable insights for the design of practical perception-aware lossy source coding systems. In particular, the proposed random coding scheme lends itself naturally to implementation via nested lattice quantization (see \cite{LHB25} for some recent development in this direction). 
The notions of universal representation, which we have clarified at a conceptual level, also merit further exploration in the context of real-world applications.

Even within the framework of the squared Wasserstein-2 perception measure, there remains a rich landscape of alternative ways to formulate the perception constraint. For instance, one may consider conditional-distribution-based formulations \cite{SCKY24} or impose marginal-distribution-based constraints with i.i.d. reconstructions \cite{XLCZ24,XLCYZ24}. Understanding the relationships among these various formulations is an intriguing open problem that warrants deeper exploration.

%
%
%
%


%

\appendices
\section{Proof of Proposition \ref{prop:inf}}\label{app:inf}

We begin with a lemma demonstrating that the MMSE estimate is preserved under weak convergence.
\begin{lemma}\label{lem:mmse}
Given $p_X$ with $\mathbb{E}[\|X\|^2]<\infty$, let $\{p^{(k)}_{X\tilde{X}}:=p_Xp^{(k)}_{\tilde{X}|X}\}_{k=1}^{\infty}$ be a sequence of probablility distributions such that $\mathbb{E}_{p^{(k)}}[X|\tilde{X}]=\tilde{X}$ $p^{(k)}$-a.s. for all $k$, and suppose that  $p^{(k)}_{X\tilde{X}}$ converges weakly to $p^*_{X\tilde{X}}$ as $k\rightarrow\infty$. Then $\mathbb{E}_{p^{*}}[X|\tilde{X}]=\tilde{X}$ $p^*$-a.s.
\end{lemma}
\begin{IEEEproof}
For $k=1,2,\ldots$, we augment $p^{(k)}_{X\tilde{X}}$ by introducing  random variables 
$Y:=[X]_M$
and $\tilde{Y}:=\mathbb{E}_{p^{(k)}}[Y|\tilde{X}]$, where $[X]_M$ denotes the clipped version of $X$ with each component restricted to the interval $[-M,M]$ and
$M>0$ is a constant. Note that $p^{(k)}_{XY}$ does not depend on $k$. 	
By the dominated convergence theorem,
\begin{align}
	\mathbb{E}[\|X-Y\|^2]=o_M(1),
\end{align}
where $o_M(1)$ denotes a quantity that tends to zero as $M\rightarrow\infty$.
The sequence $\{p^{(k)}_{XY\tilde{X}\tilde{Y}}\}_{k=1}^{\infty}$ is tight \cite[Definition in Appendix II]{GN14} since given any $\epsilon>0$,
\begin{align}
	&\mathbb{P}_{p^{(k)}}\left(\|X\|^2\leq\frac{4}{\epsilon}\mathbb{E}[\|X\|^2],\|Y\|^2\leq\frac{4}{\epsilon}\mathbb{E}[\|X\|^2],\|\tilde{X}\|^2\leq\frac{4}{\epsilon}\mathbb{E}[\|X\|^2],\|\tilde{Y}\|\leq\frac{4}{\epsilon}\mathbb{E}[\|X\|^2]\right)\nonumber\\
	&\geq 1-\mathbb{P}\left(\|X\|^2>\frac{4}{\epsilon}\mathbb{E}[\|X\|^2]\right)-\mathbb{P}\left(\|Y\|^2>\frac{4}{\epsilon}\mathbb{E}[\|X\|^2]\right)-\mathbb{P}_{p^{(k)}}\left(\|\tilde{X}\|^2>\frac{4}{\epsilon}\mathbb{E}[\|X\|^2]\right)-\mathbb{P}_{p^{(k)}}\left(\|\tilde{Y}\|^2>\frac{4}{\epsilon}\mathbb{E}[\|X\|^2]\right)\nonumber\\
	&\geq 1-\epsilon
\end{align}
for all $k$.
By Prokhorov's theorem \cite[Theorem 4]{GN14}, there exists a subsequence $\{p^{(k_m)}_{XY\tilde{X}\tilde{Y}}\}_{m=1}^{\infty}$ converging weakly to some distribution $p^*_{XY\tilde{X}\tilde{Y}}$, 
which coincides with the given $p^*_{X\tilde{X}}$ when marginalizing over $(X,\tilde{X})$.
By the property of MMSE estimation, for each $m$,
\begin{align}
	&\mathbb{E}_{p^{(k_m)}}[\|X-\tilde{Y}\|^2]=\mathbb{E}_{p^{(k_m)}}[\|X-\tilde{X}\|^2]+\mathbb{E}_{p^{(k_m)}}[\|\tilde{X}-\tilde{Y}\|^2],\\
	&\mathbb{E}_{p^{(k_m)}}[\|Y-\tilde{X}\|^2]=\mathbb{E}_{p^{(k_m)}}[\|Y-\tilde{Y}\|^2]+\mathbb{E}_{p^{(k_m)}}[\|\tilde{X}-\tilde{Y}\|^2],
\end{align}
which implies
\begin{align}
	\mathbb{E}_{p^{(k_m)}}[\|\tilde{X}-\tilde{Y}\|^2]=\frac{1}{2}\left(\mathbb{E}_{p^{(k_m)}}[\|X-\tilde{Y}\|^2]-\mathbb{E}_{p^{(k_m)}}[\|X-\tilde{X}\|^2]+\mathbb{E}_{p^{(k_m)}}[\|Y-\tilde{X}\|^2]-\mathbb{E}_{p^{(k_m)}}[\|Y-\tilde{Y}\|^2]\right).\label{eq:sum}
\end{align}
It can be verified that
\begin{align}
	&\mathbb{E}_{p^{(k_m)}}[\|X-\tilde{Y}\|^2]-\mathbb{E}_{p^{(k_m)}}[\|Y-\tilde{Y}\|^2]\nonumber\\
	&=\mathbb{E}_{p^{(k_m)}}[\|(X-Y)+(Y-\tilde{Y})\|^2]-\mathbb{E}_{p^{(k_m)}}[\|Y-\tilde{Y}\|^2]\nonumber\\
	&=\mathbb{E}[\|X-Y\|^2]+2\mathbb{E}_{p^{(k_m)}}[(X-Y)^T(Y-\tilde{Y})]\nonumber\\
	&\leq\mathbb{E}[\|X-Y\|^2]+2\sqrt{\mathbb{E}[\|X-Y\|^2]\mathbb{E}_{p^{(k_m)}}[\|Y-\tilde{Y}\|^2]}\nonumber\\
	&\leq\mathbb{E}[\|X-Y\|^2]+2\sqrt{\mathbb{E}[\|X-Y\|^2]\mathbb{E}[\|X\|^2]},\label{eq:termone}
\end{align}
and similarly, 
\begin{align}
	&\mathbb{E}_{p^{(k_m)}}[\|Y-\tilde{X}\|^2]-\mathbb{E}_{p^{(k_m)}}[\|X-\tilde{X}\|^2]\nonumber\\
	&\leq\mathbb{E}[\|X-Y\|^2]+2\sqrt{\mathbb{E}[\|X-Y\|^2]\mathbb{E}[\|X\|^2]}.\label{eq:termtwo}
\end{align}
Substituting \eqref{eq:termone} and \eqref{eq:termtwo} into \eqref{eq:sum} yields
\begin{align}
	\mathbb{E}_{p^{(k_m)}}[\|\tilde{X}-\tilde{Y}\|^2]\leq\mathbb{E}[\|X-Y\|^2]+2\sqrt{\mathbb{E}[\|X-Y\|^2]\mathbb{E}[\|X\|^2]}.
\end{align}
Since $\|\tilde{x}-\tilde{y}\|^2$ is continuous in $(\tilde{x},\tilde{y})$ and bounded from below, it follows that
\begin{align}
	\mathbb{E}_{p^*}[\|\tilde{X}-\tilde{Y}\|^2]&\leq\liminf\limits_{m\rightarrow\infty}\mathbb{E}_{p^{(k_m)}}[\|\tilde{X}-\tilde{Y}\|^2]\nonumber\\
	&\leq\mathbb{E}[\|X-Y\|^2]+2\sqrt{\mathbb{E}[\|X-Y\|^2]\mathbb{E}[\|X\|^2]}\nonumber\\
	&=o_M(1).
\end{align}
Let $\tilde{X}':=\mathbb{E}_{\mu^*}[X|\tilde{X}]$. Clearly,
\begin{align}
	\mathbb{E}_{p^*}[\|X-\tilde{X}'\|^2]\leq\mathbb{E}_{p^*}[\|X-\tilde{X}\|^2].\label{eq:onedirection}
\end{align}
Since $Y$ is supported on the bounded region $[-M,M]^L$, it follows by \cite[Theorem 3]{WV12} that
\begin{align}
	\mathbb{E}_{p^*}[\|Y-\mathbb{E}_{p^*}[Y|\tilde{Y}]\|^2]\geq\limsup\limits_{m\rightarrow\infty}\mathbb{E}_{p^{(k_m)}}[\|Y-\mathbb{E}_{p^{(k_m)}}[Y|\tilde{Y}]\|^2]=\limsup\limits_{m\rightarrow\infty}\mathbb{E}_{p^{(k_m)}}[\|Y-\tilde{Y}\|^2].\label{eq:mmse1}
\end{align}
On the other hand,  as $\|y-\tilde{y}\|^2$ is continuous in $(y,\tilde{y})$ and bounded from below, we have
\begin{align}
	\mathbb{E}_{p^*}[\|Y-\tilde{Y}\|^2]\leq\liminf\limits_{m\rightarrow\infty}\mathbb{E}_{p^{(k_m)}}[\|Y-\tilde{Y}\|^2],\label{eq:mmse2}
\end{align}
which, together with (\ref{eq:mmse1}), implies
\begin{align}
	\mathbb{E}_{p^*}[Y|\tilde{Y}]=\tilde{Y}\quad p^*-\mbox{a.s.}
\end{align} 
Therefore,
\begin{align}
	\mathbb{E}_{p^*}[\|Y-\tilde{Y}\|^2]&\leq\mathbb{E}_{p^*}[\|Y-\tilde{X}'\|^2]\nonumber\\
	&=\mathbb{E}_{p^*}[\|(Y-X)+(X-\tilde{X}')\|^2]\nonumber\\
	&=\mathbb{E}[\|X-Y\|^2]+\mathbb{E}_{p^*}[\|X-\tilde{X}'\|^2]+2\mathbb{E}_{p^*}[(Y-X)^T(X-\tilde{X}')]\nonumber\\
	&\leq\mathbb{E}[\|X-Y\|^2]+\mathbb{E}_{p^*}[\|X-\tilde{X}'\|^2]+2\sqrt{\mathbb{E}[\|X-Y\|^2]\mathbb{E}_{p^*}[\|X-\tilde{X}'\|^2]}\nonumber\\
	&\leq\mathbb{E}[\|X-Y\|^2]+\mathbb{E}_{p^*}[\|X-\tilde{X}'\|^2]+2\sqrt{\mathbb{E}[\|X-Y\|^2]\mathbb{E}[\|X\|^2]}\nonumber\\
	&=\mathbb{E}_{p^*}[\|X-\tilde{X}'\|^2]+o_M(1).\label{eq:oM2}
\end{align}
Moreover, we have
\begin{align}
	\mathbb{E}_{p^*}[\|X-\tilde{X}\|^2]&=\mathbb{E}_{p^*}[\|(X-Y)+(Y-\tilde{Y})+(\tilde{Y}-\tilde{X})\|^2]\nonumber\\
	&=\mathbb{E}[\|X-Y\|^2]+\mathbb{E}_{p^*}[\|Y-\tilde{Y}\|^2]+\mathbb{E}_{p^*}[\|\tilde{X}-\tilde{Y}\|^2]\nonumber\\
	&\quad+2\mathbb{E}_{p^*}[(X-Y)^T(Y-\tilde{Y})]+2\mathbb{E}_{p^*}[(X-Y)^T(\tilde{Y}-\tilde{X})]+2\mathbb{E}_{p^*}[(Y-\tilde{Y})^T(\tilde{Y}-\tilde{X})]\nonumber\\
	&\leq\mathbb{E}[\|X-Y\|^2]+\mathbb{E}_{p^*}[\|Y-\tilde{Y}\|^2]+\mathbb{E}_{p^*}[\|\tilde{X}-\tilde{Y}\|^2]\nonumber\\
	&\quad+2\sqrt{\mathbb{E}[\|X-Y\|^2]\mathbb{E}_{p^*}[\|Y-\tilde{Y}^2\|]}+2\sqrt{\mathbb{E}[\|X-Y\|^2]\mathbb{E}_{p^*}[\|\tilde{X}-\tilde{Y}^2\|]}\nonumber\\
	&\quad+2\sqrt{\mathbb{E}_{p^*}[\|Y-\tilde{Y}\|^2]\mathbb{E}_{p^*}[\|\tilde{X}-\tilde{Y}\|^2]}\nonumber\\
	&\leq\mathbb{E}[\|X-Y\|^2]+\mathbb{E}_{p^*}[\|Y-\tilde{Y}\|^2]+\mathbb{E}_{p^*}[\|\tilde{X}-\tilde{Y}\|^2]\nonumber\\
	&\quad+2\sqrt{\mathbb{E}[\|X-Y\|^2]\mathbb{E}[\|X\|^2]}+2\sqrt{\mathbb{E}[\|X-Y\|^2]\mathbb{E}_{p^*}[\|\tilde{X}-\tilde{Y}^2\|]}\nonumber\\
	&\quad+2\sqrt{\mathbb{E}[\|X\|^2]\mathbb{E}_{p^*}[\|\tilde{X}-\tilde{Y}\|^2]}\nonumber\\
	&=\mathbb{E}_{p^*}[\|Y-\tilde{Y}\|^2]+o_M(1).\label{eq:oM1}
\end{align}
In view of  \eqref{eq:oM2} and \eqref{eq:oM1}, it can be shown by sending $M\rightarrow\infty$ that
\begin{align}
	\mathbb{E}_{p^*}[\|X-\tilde{X}\|^2]\leq \mathbb{E}_{p^*}[\|X-\tilde{X}'\|^2],
\end{align}
which, together with \eqref{eq:onedirection}, implies
\begin{align}
	\mathbb{E}_{p^*}[\|X-\tilde{X}\|^2]=\mathbb{E}_{p^*}[\|X-\tilde{X}'\|^2].
\end{align}
Therefore, $\tilde{X}=\mathbb{E}_{p^*}[X|\tilde{X}]$ $p^*$-a.s.
\end{IEEEproof}

We are now in a position to prove Proposition \ref{prop:inf}.
For any positive integer $k$, there exist $p^{(k)}_{\tilde{X}}\in\mathcal{P}(\mathbb{R}^L)$ and 
$\mu^{(k)},\nu^{(k)}\in\Pi(p_X,p^{(k)}_{\tilde{X}})$ such that
\begin{align}
	&\mathbb{E}_{\mu^{(k)}}[X|\tilde{X}]=\tilde{X}\quad\mu^{(k)}-\mbox{a.s.},\\
	&I_{\mu^{(k)}}(X;\tilde{X})\leq R,\\
	&I_{\nu^{(k)}}(X;\tilde{X})\leq R+C,\\
	&\mathbb{E}_{\mu^{(k)}}[\|X-\tilde{X}\|^2]+\left[\left(\sqrt{\mathbb{E}_{\nu^{(k)}}[\|X-\tilde{X}\|^2]}-\sqrt{P}\right)_+\right]\leq D(R,C,P)+\frac{1}{k}.
\end{align}
The sequence $\{(\mu^{(k)},\nu^{(k)})\}_{k=1}^{\infty}$ is tight \cite[Definition in Appendix II]{GN14} since given any $\epsilon>0$,
\begin{align}
	&\max\left\{\mathbb{P}_{\mu^{(k)}}\left(\|X\|^2\leq\frac{2}{\epsilon}\mathbb{E}[\|X\|^2],\|\tilde{X}\|^2\leq\frac{2}{\epsilon}\mathbb{E}[\|X\|^2]\right),\mathbb{P}_{\nu^{(k)}}\left(\|X\|^2\leq\frac{2}{\epsilon}\mathbb{E}[\|X\|^2],\|\tilde{X}\|^2\leq\frac{2}{\epsilon}\mathbb{E}[\|X\|^2]\right)\right\}\nonumber\\
	&\geq 1-\mathbb{P}\left(\|X\|^2>\frac{2}{\epsilon}\mathbb{E}[\|X\|^2]\right)-\mathbb{P}_{p^{(k)}_{\tilde{X}}}\left(\|\tilde{X}\|^2>\frac{2}{\epsilon}\mathbb{E}[\|X\|^2]\right)\nonumber\\
	&\geq 1-\epsilon
\end{align}
for all $k$.
By Prokhorov's theorem \cite[Theorem 4]{GN14}, there exists a subsequence $\{(\mu^{(k_m)},\nu^{(k_m)})\}_{m=1}^{\infty}$ converging weakly to some distriution pair $(\mu^*,\nu^*)$. 
According to Lemma \ref{lem:mmse},
\begin{align}
	\mathbb{E}_{\mu^*}[X|\tilde{X}]=\tilde{X}\quad\mu^*-\mbox{a.s.}\label{eq:new1}
\end{align}  
Moreover, by the lower semicontinuity of mutual information in the topology of weak convergence,
\begin{align}
	&I_{\mu^*}(X;\tilde{X})\leq\liminf\limits_{m\rightarrow\infty}I_{\mu^{(k_m)}}(X;\tilde{X})\leq R,\label{eq:new2}\\
	&I_{\nu^*}(X;\tilde{X})\leq\liminf\limits_{m\rightarrow\infty}I_{\nu^{(k_m)}}(X;\tilde{X})\leq R+C.\label{eq:new3}
\end{align}
Since  $\|x-\hat{x}\|^2$ is continuous in $(x,\tilde{x})$ and bounded from below, it follows that
\begin{align}	
	&\mathbb{E}_{\mu^*}[\|X-\hat{X}\|^2]+\left[\left(\sqrt{\mathbb{E}_{\nu^*}[\|X-\tilde{X}\|^2]}-\sqrt{P}\right)_+\right]^2\nonumber\\
	&\leq\liminf\limits_{m\rightarrow\infty}\mathbb{E}_{\mu^{(k_m)}}[\|X-\hat{X}\|^2]+\left[\left(\sqrt{\mathbb{E}_{\nu^{(k_m)}}[\|X-\tilde{X}\|^2]}-\sqrt{P}\right)_+\right]^2\nonumber\\
	&\leq D(R,C,P).\label{eq:new4}
\end{align} 
In light of \eqref{eq:new1}--\eqref{eq:new4}, the infimum in (\ref{eq:inf}) is attained
at $(\mu^*,\nu^*)$. 

\section{Proof of Proposition \ref{prop:convexity}}\label{app:convexity}

The following result is needed for the proof of Proposition \ref{prop:convexity}.
\begin{lemma}\label{lem:convexity}
	The function $[(\sqrt{x}-\sqrt{y})_+]^2$ is convex in $(x,y)$ for $(x,y)\in\mathbb{R}^2_+$.
\end{lemma}
\begin{IEEEproof}
	It suffices to show that
	\begin{align}
		&[(\sqrt{(1-\lambda)x^{(0)}+\lambda x^{(1)}}-\sqrt{(1-\lambda)y^{(0)}+\lambda y^{(1)}})_+]^2\nonumber\\
		&\leq(1-\lambda)[(\sqrt{x^{(0)}}-\sqrt{y^{(0)}})_+]^2+\lambda[(\sqrt{x^{(1)}}-\sqrt{y^{(1)}})_+]^2
	\end{align}
	for any $(x^{(i)},y^{(i)})\in\mathbb{R}^2_+$, $i=0,1$, and $\lambda\in[0,1]$. We have
	\begin{align}
		&[(\sqrt{(1-\lambda)x^{(0)}+\lambda x^{(1)}}-\sqrt{(1-\lambda)y^{(0)}+\lambda y^{(1)}})_+]^2\nonumber\\
		&\leq(\sqrt{(1-\lambda)(x^{(0)}\vee y^{(0)})+\lambda (x^{(1)}\vee y^{(1)})}-\sqrt{(1-\lambda)y^{(0)}+\lambda y^{(1)}})^2\nonumber\\
		&=(1-\lambda)(x^{(0)}\vee y^{(0)})+\lambda (x^{(1)}\vee y^{(1)})+(1-\lambda)y^{(0)}+\lambda y^{(1)}\nonumber\\
		&\quad-2\sqrt{((1-\lambda)(x^{(0)}\vee y^{(0)})+\lambda (x^{(1)}\vee y^{1}))((1-\lambda)y^{(0)}+\lambda y^{(1)})}.\label{eq:suba}
	\end{align}
	Note that the function $\sqrt{xy}$ is concave in $(x,y)$ for $(x,y)\in\mathbb{R}^2_+$ since its Hessian matrix
	\begin{align}
		\left(\begin{matrix}
			-\frac{\sqrt{y}}{4x\sqrt{x}} & \frac{1}{4\sqrt{xy}}\\
			\frac{1}{4\sqrt{xy}} & -\frac{\sqrt{x}}{4y\sqrt{y}}
		\end{matrix}\right)
	\end{align}
	is negative semidefinite.
	Therefore,
	\begin{align}
		&\sqrt{((1-\lambda)(x^{(0)}\vee y^{(0)})+\lambda (x^{(1)}\vee y^{(1)}))((1-\lambda)y^{(0)}+\lambda y^{(1)})}\nonumber\\
		&\geq(1-\lambda)\sqrt{(x^{(0)}\vee y^{(0)})y^{(0)}}+\lambda\sqrt{(x^{(1)}\vee y^{(1)})y^{(1)}}. \label{eq:subb}
	\end{align}
	Substituting \eqref{eq:subb} into \eqref{eq:suba} gives
	\begin{align}
		&[(\sqrt{(1-\lambda)x^{(0)}+\lambda x^{(1)}}-\sqrt{(1-\lambda)y^{(0)}+\lambda y^{(1)}})_+]^2\nonumber\\
		&\leq(1-\lambda)(x^{(0)}\vee y^{(0)})+\lambda (x^{(1)}\vee y^{(1)})+(1-\lambda)y^{(0)}+\lambda y^{(1)}-2(1-\lambda)\sqrt{(x^{(0)}\vee y^{(0)})y^{(0)}}-2\lambda\sqrt{(x^{(1)}\vee y^{(1)})y^{(1)}}\nonumber\\
		&=(1-\lambda)(\sqrt{x^{(0)}\vee y^{(0)}}-\sqrt{y^{(0)}})^2+\lambda(\sqrt{x^{(1)}\vee y^{(1)}}-\sqrt{y^{(1)}})^2\nonumber\\
		&=(1-\lambda)[(\sqrt{x^{(0)}}-\sqrt{y^{(0)}})_+]^2+\lambda[(\sqrt{x^{(1)}}-\sqrt{y^{(1)}})_+]^2,
	\end{align}
	establishing the desired result.
\end{IEEEproof}

Now we proceed to prove Proposition \ref{prop:convexity}. The function $D(R,C,P)$ is clearly decreasing in $(R,C,P)$. So it suffices to prove its convexity and continuity. By Proposition \ref{prop:inf}, for $i=0,1$, given a triple $(R^{(i)},C^{(i)},P^{(i)})$, there exist $\mu^{(i)},\nu^{(i)}\in\Pi(p_X,p_{\tilde{X}^{(i)}})$ such that
\begin{align}
	&\mathbb{E}_{\mu^{(i)}}[\|X-\tilde{X}^{(i)}\|^2]+\left[\left(\sqrt{\mathbb{E}_{\nu^{(i)}}[\|X-\tilde{X}^{(i)}\|^2]}-\sqrt{P^{(i)}}\right)_+\right]^2= D(R^{(i)},C^{(i)},P^{(i)}),\\
	&\mathbb{E}_{\mu^{(i)}}[X|\tilde{X}^{(i)}]=\hat{X}^{(i)}\quad\mu^{(i)}-\mbox{a.s.},\\
	&I_{\mu^{(i)}}(X;\tilde{X}^{(i)})\leq R^{(i)},\\
	&I_{\nu^{(i)}}(X;\tilde{X}^{(i)})\leq R^{(i)}+C^{(i)}.
\end{align}
 Let $(X,\tilde{X})\sim\mu^{(\lambda)}:=(1-\lambda)\mu^{(0)}+\lambda\mu^{(1)}$ or $\sim\nu^{(\lambda)}:=(1-\lambda)\nu^{(0)}+\lambda\nu^{(1)}$, $R^{(\lambda)}:=(1-\lambda)R^{(0)}+\lambda R^{(1)}$, $C^{(\lambda)}:=(1-\lambda)C^{(0)}+\lambda C^{(1)}$, and $P^{(\lambda)}:=(1-\lambda)P^{(0)}+\lambda P^{(1)}$, where $\lambda\in[0,1]$. It can be verified that
\begin{align}
	&\mathbb{E}_{\mu^{(\lambda)}}[X|\tilde{X}^{(\lambda)}]=\tilde{X}^{(\lambda)}\quad\mu^{(\lambda)}-\mbox{a.s.},\\
	&I_{\mu^{(\lambda)}}(X;\tilde{X}^{(\lambda)})\leq(1-\lambda)I_{\mu^{(0)}}(X;\tilde{X}^{(0)})+\lambda I_{\mu^{(1)}}(X;\tilde{X}^{(1)})\leq  R^{(\lambda)},\\		&I_{\nu^{(\lambda)}}(X;\tilde{X}^{(\lambda)})\leq(1-\lambda)I_{\nu^{(0)}}(X;\tilde{X}^{(0)})+\lambda I_{\nu^{(1)}}(X;\tilde{X}^{(1)})\leq R^{(\lambda)}+C^{(\lambda)},
\end{align}
which, together with the fact $\mu^{(\lambda)},\nu^{(\lambda)}\in\Pi(p_X,p_{\tilde{X}^{(\lambda)}})$, implies
\begin{align}
	D(R^{(\lambda)}, C^{(\lambda)}, P^{(\lambda)})\leq\mathbb{E}_{\mu^{(\lambda)}}[\|X-\tilde{X}^{(\lambda)}\|^2]+\left[\left(\sqrt{\mathbb{E}_{\nu^{(\lambda)}}[\|X-\tilde{X}^{(\lambda)}\|^2]}-\sqrt{P^{(\lambda)}}\right)_+\right]^2.\label{eq:convexitybound}
\end{align}
Moreover, by Lemma \ref{lem:convexity}, we have
\begin{align}
	&\mathbb{E}_{\mu^{(\lambda)}}[\|X-\tilde{X}^{(\lambda)}\|^2]+\left[\left(\sqrt{\mathbb{E}_{\nu^{(\lambda)}}[\|X-\tilde{X}^{(\lambda)}\|^2]}-\sqrt{ P^{(\lambda)}}\right)_+\right]^2\nonumber\\
	&\leq(1-\lambda)\left(\mathbb{E}_{\mu^{(0)}}[\|X-\tilde{X}^{(0)}\|^2]+\left[\left(\sqrt{\mathbb{E}_{\nu^{(0)}}[\|X-\tilde{X}^{(0)}\|^2]}-\sqrt{P^{(0)}}\right)_+\right]^2\right)\nonumber\\
	&\quad +\lambda\left(\mathbb{E}_{\mu^{(1)}}[\|X-\tilde{X}^{(1)}\|^2]+\left[\left(\sqrt{\mathbb{E}_{\nu^{(1)}}[\|X-\tilde{X}^{(1)}\|^2]}-\sqrt{P^{(1)}}\right)_+\right]^2\right)\nonumber\\
	&\leq(1-\lambda)D(R^{(0)},C^{(0)},P^{(0)})+\lambda D(R^{(1)},C^{(1)},P^{(1)}).\label{eq:convexityeps}
\end{align}
Combining \eqref{eq:convexitybound}  and \eqref{eq:convexityeps} proves the convexity of $D(R,C,P)$.

The convexity property ensures that 
the function $D(R,C,P)$ is continuous in $(R,C,P)$ over the interior of its domain. So it remains to show that this continuity extends to the boundary points. The weak convergence argument used in the proof of Proposition \ref{prop:inf} can be easily leveraged to estalsih  the lower semicontinuity of
$D(R,C,P)$, i.e., 
\begin{align}
	\liminf\limits_{(R',C',P')\rightarrow(R,C,P)}D(R',C',P')\geq D(R,C,P).\label{eq:liminf}
\end{align}
Consider the boundary case where $R=0$, $C>0$, and $P>0$. 
We have
\begin{align}
\limsup\limits_{(R',C',P')\rightarrow(0,C,P)}D(R',C',P')&\leq\limsup\limits_{(C',P')\rightarrow(C,P)}D(0,C',P')\\
&=D(0,C,P),\label{eq:limsup}
\end{align}
where the inequality follows by the fact that $D(R',C',P')$ is decreasing in $R'$ for fixed $(C',P')$, while the equality holds due to the continuity of $D(0,C,P)$ in $(C,P)$ for $C>0$ and $P>0$, as implied by its convexity. In view of \eqref{eq:liminf} and \eqref{eq:limsup},
\begin{align}
	\lim\limits_{(R',C',P')\rightarrow(0,C,P)}D(R',C',P')=D(0,C,P).
\end{align}
Other boundary cases can be treated similarly. This completes the proof of Proposition \ref{prop:convexity}.

\section{Proof of Theorem \ref{thm:main}}

\label{app:main}

The proof is divided into two parts: achievability and converse. 

\subsection{Achievability}

The
achievability part is dedicated to proving that $D^{*}(R,C,P)\leq D(R,C,P)$. When $R=0$, we have $\tilde{X}=\mathbb{E}[X]$ a.s., and thus 
\begin{align}
D(0,C,P)=\mathbb{E}[\|X-\mathbb{E}[X]\|^{2}]+\left[\left(\sqrt{\mathbb{E}[\|X-\mathbb{E}[X]\|^{2}]}-\sqrt{P}\right)_{+}\right]^{2},
\end{align}
which is clearly achievable using the interpolation scheme described
in Section \ref{subsec:nC}. 
So it remains to consider the case where $R>0$. Given $\epsilon<R$,
let $\mu$ and $\nu$ be two coupligs in $\Pi(p_{X},p_{\tilde{X}})$
for some $p_{\tilde{X}}\in\mathcal{P}(\mathbb{R}^{L})$ such that
\begin{align}
 & \mathbb{E}_{\mu}[\|X-\tilde{X}\|^{2}]+\left[\left(\sqrt{\mathbb{E}_{\mu}[\|X-\tilde{X}\|^{2}]}-\sqrt{P}\right)_{+}\right]\leq D(R-\epsilon,C,P),\\
 & \mathbb{E}_{\mu}[X|\tilde{X}]=\tilde{X}\quad\mu-\mbox{a.s.},\\
 & I_{\mu}(X;\tilde{X})\leq R-\epsilon,\\
 & I_{\nu}(X;\tilde{X})\leq R+C-\epsilon.
\end{align}
The crux of the proof lies in establishing the existence of an encoder with the following properties.
\begin{align}
 & \frac{1}{n}\log|\mathcal{J}|\leq R,\label{eq:lei1}\\
 & \frac{1}{n}\log|\mathcal{K}|\leq C,\label{eq:lei2}\\
 & \frac{1}{n}\mathbb{E}[\|X^{n}-\tilde{X}^{n}\|^{2}]\leq\mathbb{E}_{\mu}[\|X-\tilde{X}\|^{2}]+\epsilon,\label{eq:lei3}\\
 & \frac{1}{n}W_{2}^{2}(p_{X^{n}},p_{\tilde{X}^{n}})\leq\mathbb{E}_{\nu}[\|X-\tilde{X}\|^{2}]+\epsilon,\label{eq:lei4}
\end{align}
where $\tilde{X}^{n}:=\mathbb{E}[X^{n}|J,K]$. In light of \eqref{eq:nDRP},
the end-to-end distortion achieved by such an encoder, combined with
the optimal decoder based on the interpolation scheme, can be bounded
as 
\begin{align}
\frac{1}{n}\mathbb{E}[\|X^{n}-\hat{X}^{n}\|^{2}] & =\frac{1}{n}\mathbb{E}[\|X^{n}-\tilde{X}^{n}\|^{2}]+\left[\left(\frac{1}{\sqrt{n}}W_{2}(p_{X^{n}},p_{\tilde{X}^{n}})-\sqrt{P}\right)_{+}\right]^{2}\nonumber \\
 & \leq\mathbb{E}_{\mu}[\|X-\tilde{X}\|^{2}]+\epsilon+\left[\left(\sqrt{\mathbb{E}_{\nu}[\|X-\tilde{X}\|^{2}]+\epsilon}-\sqrt{P}\right)_{+}\right]^{2}\nonumber \\
 & \leq\mathbb{E}_{\mu}[\|X-\tilde{X}\|^{2}]+\epsilon+\left[\left(\sqrt{\mathbb{E}_{\nu}[\|X-\tilde{X}\|^{2}]}-\sqrt{P}\right)_{+}+\sqrt{\epsilon}\right]^{2}\nonumber \\
 & \leq\mathbb{E}_{\mu}[\|X-\tilde{X}\|^{2}]+\left[\left(\sqrt{\mathbb{E}_{\nu}[\|X-\tilde{X}\|^{2}]}-\sqrt{P}\right)_{+}\right]^{2}+2\epsilon+2\sqrt{\epsilon\mathbb{E}_{\nu}[\|X-\tilde{X}\|^{2}]}\nonumber \\
 & \leq\mathbb{E}_{\mu}[\|X-\tilde{X}\|^{2}]+\left[\left(\sqrt{\mathbb{E}_{\nu}[\|X-\tilde{X}\|^{2}]}-\sqrt{P}\right)_{+}\right]^{2}+2\epsilon+4\sqrt{\epsilon\mathbb{E}[\|X\|^{2}]}\nonumber \\
 & \leq D(R-\epsilon,C,P)+2\epsilon+4\sqrt{\epsilon\mathbb{E}[\|X\|^{2}]}.
\end{align}
This yields 
\begin{align}
D^{*}(R,C,P)\leq D(R-\epsilon,C,P)+2\epsilon+4\sqrt{\epsilon\mathbb{E}[\|X\|^{2}]}.
\end{align}
Sending $\epsilon\rightarrow0$ and invoking Proposition \ref{prop:convexity}
(more specifically, the continuity of $D(R,C,P)$ in $R$ for fixed
$(C,P)$) establishes $D^{*}(R,C,P)\leq D(R,C,P)$.

\textbf{Step 1 (Constructing a random code): } We begin by constructing a random code that relies on additional randomness, which will be removed in a subsequent step.

The construction proceeds as follows.  We generate\footnote{For notational simplicity, we assume that $2^{nR}$ and $2^{nC}$
	are integers. Otherwise, they can be replaced by their floor values.} a codebook
$\mathcal{C}=\{\tilde{X}^{n}(j,k):j\in[2^{nR}],k\in[2^{nC}]\}$, where
the codewords $\tilde{X}^{n}(j,k)$ are drawn independently according
to the same product distribution $p_{\tilde{X}}^{n}$. Partition $\mathcal{C}$
into $2^{nC}$ subcodebooks $\mathcal{C}(k)=\{\tilde{X}^{n}(j,k):j\in[2^{nR}]\}$.
 We randomly and uniformly select a codeword from $\mathcal{C}$
and pass it through the memoryless channel $\mu_{X|\tilde{X}}^{n}$, resulting in the output $X^{\prime n}$. 
Let $J\sim p_{J}:=\mathrm{Unif}[2^{nR}]$ and $K\sim p_{K}:=\mathrm{Unif}[2^{nC}]$
be two independent random variables. The induced joint distribution
is given by
\begin{align}
p_{JK\tilde{X}^{n}X^{\prime n}|\mathcal{C}}=2^{-n(R+C)}\delta_{\tilde{X}^{n}(J,K)}\mu_{X|\tilde{X}}^{n},
\end{align}
where  $\delta_{x}$
denotes the Dirac measure at $x$.
This joint distribution implies that
\begin{align}
p_{X^{\prime n}|K\mathcal{C}}=2^{-nR}\sum_{j\in[2^{nR}]}p_{X|\tilde{X}}^{n}(\cdot|\tilde{X}^{n}(j,K)).
\end{align}

We now introduce a soft-covering lemma with respect to the Wasserstein-2 distance. The proof is deferred to Appendix \ref{subsec:Proof-of-Lemma}.
\begin{lemma}
\label{lem:soft-covering} Let $p_{XY}$ be a joint distribution
defined on $\mathbb{R}^{L}\times\mathbb{R}^{L}$ such that $\mathbb{E}[\|Y\|^{2}]<\infty$. Moreover, let
\begin{align}
	p_{JX^{n}Y^{n}|\mathcal{C}}:=p_{J}\delta_{X^{n}(J)}p_{Y|X}^{n}
\end{align}
where $\mathcal{C}:=\{X^{n}(j):j\in[2^{nR}]\}$ with each $X^{n}(j)$ drawn
independently according to the same product distribution $p_{X}^{n}$ and
$J\sim p_{J}:=\mathrm{Unif}[2^{nR}]$. 
If $R>I(X;Y)$, then
\begin{align}
	\lim\limits_{n\rightarrow\infty}\frac{1}{n}\mathbb{E}_{\mathcal{C}}[W_{2}^{2}(p_{Y^{n}|\mathcal{C}},p_{Y}^{n})]=0.
\end{align}
\end{lemma}

By Lemma \ref{lem:soft-covering},  
\begin{equation}
\lim\limits_{n\rightarrow\infty}\frac{1}{n}\mathbb{E}_{(K,\mathcal{C})}\left[W_{2}^{2}(p_{X^{\prime n}|K\mathcal{C}},p_{X}^{n})\right]=0.\label{eq:-5}
\end{equation}
For each $(k,c)$, denote the coupling that attains $W_{2}^{2}(p_{X^{\prime n}|K=k,\mathcal{C}=c},p_{X}^{n})$
 by $p_{X^{\prime n}X^{n}|K=k,\mathcal{C}=c}$. 
In light of measurable selection
theorems (see, e.g., \cite[Proposition 7.50]{bertsekas1996stochastic}),  
 $p_{X^{\prime n}X^{n}|K\mathcal{C}}$ can be assumed to be a
regular conditional distribution.
Define 
\begin{equation}
p_{JK\tilde{X}^{n}X^{\prime n}X^{n}|\mathcal{C}}:=p_{JK\tilde{X}^{n}X^{\prime n}|\mathcal{C}}p_{X^{n}|X^{\prime n}K\mathcal{C}}.\label{eq:-4}
\end{equation}
Under this joint distribution, we have $J\sim\mathrm{Unif}[2^{nR}]$,
$K\sim\mathrm{Unif}[2^{nC}]$, and $p_{X^{n}|K\mathcal{C}}=p_{X}^{n}$.
By treating $p_{J|X^{n}K\mathcal{C}}$  as the encoder and $(j,k)\mapsto\tilde{X}^{n}(j,k)$
as the decoder, the resulting distribution is exactly
$p_{JK\tilde{X}^{n}X^{n}|\mathcal{C}}$, which is the marginal of the joint
distribution in \eqref{eq:-4}.

\textbf{Step 2 (Evaluating the performance): } We now evaluate the
performance of this random code. It is easy to see that
\eqref{eq:lei1} and \eqref{eq:lei2} are satisfied. Moreover, by the Cauchy--Schwarz inequality,
\begin{align*}
\mathbb{E}_{(p,\mathcal{C})}[\|X^{n}-\tilde{X}^{n}\|^{2}] & =\mathbb{E}_{(p,\mathcal{C})}[\|X^{n}-X^{\prime n}+X^{\prime n}-\tilde{X}^{n}\|^{2}]\\
 & \le\mathbb{E}_{(p,\mathcal{C})}[\|X^{n}-X^{\prime n}\|^{2}]+\mathbb{E}_{(p,\mathcal{C})}[\|X^{\prime n}-\tilde{X}^{n}\|^{2}]+2\sqrt{\mathbb{E}_{(p,\mathcal{C})}[\|X^{n}-X^{\prime n}\|^{2}]\mathbb{E}_{(p,\mathcal{C})}[\|X^{\prime n}-\tilde{X}^{n}\|^{2}]}\\
 & =\mathbb{E}_{(K,\mathcal{C})}[W_{2}^{2}(p_{X^{\prime n}|K\mathcal{C}},p_{X}^{n})]+n\mathbb{E}_{\mu}[(X-\tilde{X})^{2}]+2\sqrt{n\mathbb{E}_{(K,\mathcal{C})}[W_{2}^{2}(p_{X^{\prime n}|K\mathcal{C}},p_{X}^{n})]\mathbb{E}_{\mu}[(X-\tilde{X})^{2}]},
\end{align*}
which, together with \eqref{eq:-5}, yields 
\begin{align}
\limsup_{n\to\infty}\frac{1}{n}\mathbb{E}_{(p,\mathcal{C})}[\|X^{n}-\tilde{X}^{n}\|^{2}] & \le\mathbb{E}_{\mu}[(X-\tilde{X})^{2}].\label{eq:}
\end{align}
Therefore, \eqref{eq:lei3} holds for sufficiently large $n$.

It remains to verify \eqref{eq:lei4}.
To this end, it suffices to show that there is a coupling $q_{X^{n}\tilde{X}^{n}|\mathcal{C}}$
of $p_{X}^{n}$ and $p_{\tilde{X}^{n}|\mathcal{C}}$ such that
\begin{align}
\frac{1}{n}\mathbb{E}_{(q,\mathcal{C})}[\|X^{n}-\tilde{X}^{n}\|^{2}]\leq\mathbb{E}_{\nu}[\|X-\tilde{X}\|^{2}]+\epsilon.
\end{align}
We construct the desired coupling using a scheme analogous to the one described above. Specifically,
we input the codeword $\tilde{X}^{n}(J,K)$, selected  from the codebook $\mathcal{C}$,
into the memoryless channel $\nu_{X|\tilde{X}}^{n}$, and let $X^{\prime n}$ denote the corresponding channel output. The induced
joint distribution is given by
\begin{align}
q_{JK\tilde{X}^{n}X^{\prime n}|\mathcal{C}}=2^{-n(R+C)}\delta_{\tilde{X}^{n}(J,K)}\nu_{X|\tilde{X}}^{n}.
\end{align}
Since $I_{\nu}(X;\tilde{X})\leq R+C-\epsilon$, it follows by
Lemma \ref{lem:soft-covering} that
\begin{align}
\lim\limits_{n\rightarrow\infty}\frac{1}{n}\mathbb{E}_{(K,\mathcal{C})}\left[W_{2}^{2}(q_{X^{\prime n}|\mathcal{C}},p_{X}^{n})\right]=0.
\end{align}
Let $q_{X^{\prime n}X^{n}|\mathcal{C}}$ denote the coupling that attains
$W_{2}^{2}(q_{X^{\prime n}|\mathcal{C}},p_{X}^{n})$,
and  define
\begin{align}
q_{JK\tilde{X}^{n}X^{\prime n}X^{n}|\mathcal{C}}:=q_{JK\tilde{X}^{n}X^{\prime n}|\mathcal{C}}q_{X^{n}|X^{\prime n}\mathcal{C}}.
\end{align}
Under this joint distribution, we have $q_{JK\tilde{X}^{n}|\mathcal{C}}=p_{JK\tilde{X}^{n}|\mathcal{C}}$
and $q_{X^{n}|\mathcal{C}}=p_{X}^{n}$.  Similar to \eqref{eq:},
\begin{align}
\limsup_{n\to\infty}\frac{1}{n}\mathbb{E}_{(q,\mathcal{C})}[\|X^{n}-\tilde{X}^{n}\|^{2}] & \le\mathbb{E}_{\nu}[(X-\tilde{X})^{2}].\label{eq:similar}
\end{align}
Therefore, \eqref{eq:lei4} holds for sufficiently large $n$. 

Note that the random code constructed above does not necessarily satisfy $\tilde{X}^{n}=\mathbb{E}_{p}[X^{n}|J,K,\mathcal{C}]$
 a.s. We now modify the scheme by replacing $\tilde{X}^{n}$ with $\mathbb{E}_{p}[X^{n}|J,K,\mathcal{C}]$. Due to the optimality  of the MMSE estimate, 
 \eqref{eq:lei3} remains valid under this substitution.
Furthermore,  noting that $\mathbb{E}_{p}[X^{\prime n}|J,K,\mathcal{C}]=\tilde{X}^{n}(J,K)$,
we have 
\begin{align}
\frac{1}{n}\mathbb{E}_{(p,\mathcal{C})}[\|\tilde{X}^{n}-\mathbb{E}_{p}[X^{n}|J,K,\mathcal{C}]\|^{2}] & =\frac{1}{n}\mathbb{E}_{(p,\mathcal{C})}[\|\mathbb{E}_{p}[X^{\prime n}|J,K,\mathcal{C}]-\mathbb{E}_{p}[X^{n}|J,K,\mathcal{C}]\|^{2}]\nonumber \\
 & \le\frac{1}{n}\mathbb{E}_{(p,\mathcal{C})}[\|X^{\prime n}-X^{n}\|^{2}]\nonumber \\
 & =\frac{1}{n}\mathbb{E}_{(K,\mathcal{C})}[W_{2}^{2}(p_{X^{\prime n}|K\mathcal{C}},p_{X}^{n})]\nonumber \\
 & \to 0,\label{eq:-6}
\end{align}
as $n\rightarrow\infty$, where the second step follows from Jensen's inequality, and the last
step is due to \eqref{eq:-5}. Invoking the Cauchy-Schwarz inequality gives 
\begin{align}
 & \mathbb{E}_{(q,\mathcal{C})}[\|X^{n}-\mathbb{E}_{p}[X^{n}|J,K,\mathcal{C}]\|^{2}]\nonumber\\ & =\mathbb{E}_{(q,\mathcal{C})}[\|X^{n}-\tilde{X}^{n}+\tilde{X}^{n}-\mathbb{E}_{p}[X^{n}|J,K,\mathcal{C}]\|^{2}]\nonumber\\
 & \le\mathbb{E}_{(q,\mathcal{C})}[\|X^{n}-\tilde{X}^{n}\|^{2}]+\mathbb{E}_{(q,\mathcal{C})}[\|\tilde{X}^{n}-\mathbb{E}_{p}[X^{n}|J,K,\mathcal{C}]\|^{2}]+2\sqrt{\mathbb{E}_{(q,\mathcal{C})}[\|X^{n}-\tilde{X}^{n}\|^{2}]\mathbb{E}_{(q,\mathcal{C})}[\|\tilde{X}^{n}-\mathbb{E}_{p}[X^{n}|J,K,\mathcal{C}]\|^{2}]}.\label{eq:yield}
\end{align}
Since $q_{JK\tilde{X}^{n}|\mathcal{C}}=p_{JK\tilde{X}^{n}|\mathcal{C}}$, it follows by \eqref{eq:-6} that
\begin{align} \lim\limits_{n\rightarrow\infty}\frac{1}{n}\mathbb{E}_{(q,\mathcal{C})}[\|\tilde{X}^{n}-\mathbb{E}_{p}[X^{n}|J,K,\mathcal{C}]\|^{2}]=\lim\limits_{n\rightarrow\infty}\frac{1}{n}\mathbb{E}_{(p,\mathcal{C})}[\|\tilde{X}^{n}-\mathbb{E}_{p}[X^{n}|J,K,\mathcal{C}]\|^{2}]=0,
\end{align}
which, together with \eqref{eq:yield} and \eqref{eq:similar}, yields
\begin{align}
\limsup_{n\to\infty}\frac{1}{n}\mathbb{E}_{(q,\mathcal{C})}[\|X^{n}-\mathbb{E}_{p}[X^{n}|J,K,\mathcal{C}]\|^{2}] & =\limsup_{n\to\infty}\frac{1}{n}\mathbb{E}_{(q,\mathcal{C})}[\|X^{n}-\tilde{X}^{n}\|^{2}]\nonumber\\
 & \le\mathbb{E}_{\nu}[(X-\hat{X})^{2}].
\end{align}

\textbf{Step 3 (Derandomization): }In the coding scheme  described above, the construction of the codebook $\mathcal{C}$ requires additional common randomness beyond the shared random seed $K$. We now proceed to eliminate this additional randomness. By the support
lemma, there exist two realizations $c_{0}$ and $c_{1}$ of the codebook $\mathcal{C}$, along with a scalar
a number $\lambda\in[0,1]$, such that 
\begin{align*}
\mathbb{E}_{(p,\mathcal{C})}\left[f(\mathcal{C})\right] & =(1-\lambda)f(c_{0})+\lambda f(c_{1}),\\
\mathbb{E}_{(q,\mathcal{C})}\left[g(\mathcal{C})\right] & =(1-\lambda)g(c_{0})+\lambda g(c_{1}),
\end{align*}
where $f(c):=\mathbb{E}_{p}\left[\|X^{n}-\mathbb{E}_{p}[X^{n}|J,K,\mathcal{C}]\|^{2}|\mathcal{C}=c\right]$
and $g(c):=\mathbb{E}_{q}\left[\|X^{n}-\mathbb{E}_{p}[X^{n}|J,K,\mathcal{C}]\|^{2}|\mathcal{C}=c\right]$.
By employing a time-sharing strategy that uses $c_0$ for a proportion $1-\lambda$  of the time and 
$c_1$ for a proportion $\lambda$ of the time, we conclude that \eqref{eq:lei1}--\eqref{eq:lei4} are satisfied by a deterministic code.

\subsection{Converse }

We now turn to the converse, aiming to show that $D^{*}(R,C,P)\geq D(R,C,P)$.
Consider an arbitrary length-$n$ perception-aware lossy source coding
system satisfying 
\begin{align}
 & \frac{1}{n}\log|\mathcal{J}|\leq R,\\
 & \frac{1}{n}\log|\mathcal{K}|\leq C,\\
 & \frac{1}{n}W_{2}^{2}(p_{X^{n}},p_{\hat{X}^{n}})\leq P,
\end{align}
and denote its induced distribution by $p_{X^{n}JK\hat{X}^{n}}$.
We augment $p_{X^{n}JK\hat{X}^{n}}$ by introducing a random variable
$T$, which is uniformly distributed over $\{1,2,\ldots,n\}$ and
independent of $(X^{n},J,K,\hat{X}^{n})$ under $p_{X^{n}JK\hat{X}^{n}T}$
Let $\tilde{X}^{n}:=\mathbb{E}_{p}[X^{n}|J,K]$. Note that 
\begin{align}
R & \geq\frac{1}{n}\log|\mathcal{J}|\nonumber \\
 & \geq\frac{1}{n}H_{p}(J)\nonumber \\
 & \geq\frac{1}{n}I_{p}(X^{n};J|K)\nonumber \\
 & =\frac{1}{n}I_{p}(X^{n};J,K)\nonumber \\
 & \geq\frac{1}{n}I_{p}(X^{n};\tilde{X}^{n})\nonumber \\
 & =\frac{1}{n}\sum\limits _{t=1}^{n}I_{p}(X(t);\tilde{X}^{n}|X^{t-1})\nonumber \\
 & =\frac{1}{n}\sum\limits _{t=1}^{n}I_{p}(X(t);\tilde{X}^{n},X^{t-1})\nonumber \\
 & \geq\frac{1}{n}\sum\limits _{t=1}^{n}I_{p}(X(t);\tilde{X}(t))\nonumber \\
 & =I_{p}(X(T);\tilde{X}(T)|T)\nonumber \\
 & =I_{p}(X(T);\tilde{X}(T),T)\nonumber \\
 & \geq I_{p}(X(T);\tilde{X}(T)).\label{eq:comb1}
\end{align}
Moreover, 
\begin{align}
\mathbb{E}_{p}[X(T)|\tilde{X}(T)] & =\mathbb{E}_{p}[\mathbb{E}_{p}[X(T)|\tilde{X}(T),T]|\tilde{X}(T)]\nonumber \\
 & =\mathbb{E}_{p}[\tilde{X}(T)|\tilde{X}(T)]\nonumber \\
 & =\tilde{X}(T)\quad p-\mbox{a.s.}\label{eq:timesharing}
\end{align}
In light of \eqref{eq:nDRP}, 
\begin{align}
\frac{1}{n}\mathbb{E}_{p}[\|X^{n}-\hat{X}^{n}\|^{2}]\geq\frac{1}{n}\mathbb{E}_{p}[\|X^{n}-\tilde{X}^{n}\|^{2}]+\left[\left(\frac{1}{\sqrt{n}}W_{2}(p_{X^{n}},p_{\tilde{X}^{n}})-\sqrt{P}\right)_{+}\right]^{2}.\label{eq:endtoend}
\end{align}
We have 
\begin{align}
\frac{1}{n}\mathbb{E}_{p}[\|X^{n}-\tilde{X}^{n}\|^{2}] & =\mathbb{E}_{p}[\mathbb{E}_{p}[\|X(T)-\tilde{X}(T)\|^{2}|T]]\nonumber \\
 & =\mathbb{E}_{p}[\|X(T)-\tilde{X}(T)\|^{2}].\label{eq:T1}
\end{align}
Let $q_{X^{n}\tilde{X}^{n}}$ denote the coupling of $p_{X^{n}}$
and $p_{\tilde{X}^{n}}$ that achieves $W_{2}^{2}(p_{X^{n}},p_{\tilde{X}^{n}})$.
We augment $q_{X^{n}\tilde{X}^{n}}$ by introducing a random variable
$T$, which is uniformly distributed over $\{1,2,\ldots,n\}$ and
independent of $(X^{n},\tilde{X}^{n})$ under $q_{X^{n}\tilde{X}^{n}T}$.
It can be verified that 
\begin{align}
\frac{1}{n}W_{2}^{2}(p_{X^{n}},p_{\tilde{X}^{n}}) & =\frac{1}{n}\mathbb{E}_{q}[\|X^{n}-\tilde{X}^{n}\|^{2}]\nonumber \\
 & =\mathbb{E}_{q}[\mathbb{E}_{q}[\|X(T)-\tilde{X}(T)\|^{2}|T]]\nonumber \\
 & =\mathbb{E}_{q}[\|X(T)-\tilde{X}(T)\|^{2}].\label{eq:T2}
\end{align}
Substituting \eqref{eq:T1} and \eqref{eq:T2} into \eqref{eq:endtoend}
yields 
\begin{align}
\frac{1}{n}\mathbb{E}_{p}[\|X^{n}-\hat{X}^{n}\|^{2}]\geq\mathbb{E}_{p}[\|X(T)-\tilde{X}(T)\|^{2}]+\left[\left(\sqrt{\mathbb{E}_{q}[\|X(T)-\tilde{X}(T)\|^{2}]}-\sqrt{P}\right)_{+}\right]^{2}.\label{eq:comb3}
\end{align}
In addition, 
\begin{align}
R+C & \geq\frac{1}{n}\log|\mathcal{J}|+\frac{1}{n}\log|\mathcal{K}|\nonumber \\
 & \geq\frac{1}{n}H_{p}(J,K)\nonumber \\
 & \geq\frac{1}{n}H_{p}(\tilde{X}^{n})\nonumber \\
 & =\frac{1}{n}H_{q}(\tilde{X}^{n})\nonumber \\
 & \geq\frac{1}{n}I_{q}(X^{n};\tilde{X}^{n})\nonumber \\
 & =\frac{1}{n}\sum\limits _{t=1}^{n}I_{q}(X(t);\tilde{X}^{n}|X^{t-1})\nonumber \\
 & =\frac{1}{n}\sum\limits _{t=1}^{n}I_{q}(X(t);\tilde{X}^{n},X^{t-1})\nonumber \\
 & \geq\frac{1}{n}\sum\limits _{t=1}^{n}I_{q}(X(t);\tilde{X}(t))\nonumber \\
 & =I_{q}(X(T);\tilde{X}(T)|T)\nonumber \\
 & =I_{q}(X(T);\tilde{X}(T),T)\nonumber \\
 & \geq I_{q}(X(T);\tilde{X}(T)).\label{eq:R+C}
\end{align}
Note that $p_{X(T)}=q_{X(T)}=p_{X}$ and $p_{\tilde{X}(T)}=q_{\tilde{X}(T)}$.
In view of \eqref{eq:comb1}, \eqref{eq:timesharing}, \eqref{eq:comb3},
and \eqref{eq:R+C}, one can readily complete the proof by setting $\mu:=p_{X(T)\tilde{X}(T)}$
and $\nu:=q_{X(T)\tilde{X}(T)}$.

\subsection{\label{subsec:Proof-of-Lemma}Proof of Lemma \ref{lem:soft-covering}}

Given
$\mathcal{C}=c$, we write $p_{Y^{n}|\mathcal{C}=c}$ as $q_{Y^{n}}$ for
brevity. Let $q_{Y^n}(\cdot|B_r)$ and $p^n_{Y}(\cdot|B_r)$ denote the conditional probability measures induced by $q_{Y^n}$ and $p^n_{Y}$, respectively, given that $Y^n\in B_r$, where $B_{r}$ is the Euclidean ball in $\mathbb{R}^{nL}$  centered
at the origin, with radius $r$ to be specified later.
Moreover, let $p_{Y^{\prime n}Y^{n}}$ be a maximal coupling of $q_{Y^{n}}(\cdot|B_{r})$
and $p_{Y}^{n}(\cdot|B_{r})$, i.e., a coupling for which $\mathbb{P}_{p}(Y^{\prime n}\neq Y^{n})$ equals the total variation distance $\|q_{Y^{n}}(\cdot|B_{r})-p_{Y}^{n}(\cdot|B_{r})\|_{\mathrm{TV}}$. Based on $p_{Y^{\prime n}Y^{n}}$,
we construct a coupling $\hat{p}_{Y^{\prime n}Y^{n}}$ of
$q_{Y^{n}}$ and $p_{Y}^{n}$  as follows: 
\begin{align}
\hat{p}_{Y^{\prime n}Y^{n}} :=abp_{Y^{\prime n}Y^{n}}+(1-a)bq_{Y^{n}}(\cdot|B_{r}^{c})p_{Y}^{n}(\cdot|B_{r})+a(1-b)q_{Y^{n}}(\cdot|B_{r})p_{Y}^{n}(\cdot|B_{r}^{c})+(1-a)(1-b)q_{Y^{n}}(\cdot|B_{r}^{c})p_{Y}^{n}(\cdot|B_{r}^{c}),
\end{align}
where $a:=q_{Y^{n}}(B_{r})$ and $b=p_{Y}^{n}(B_{r})$.  It can be shown by invoking the
inequality $\|y^{\prime n}-y^{n}\|^{2}\le2(\|y^{\prime n}\|^{2}+\|y^{n}\|^{2})$ that 
\begin{align}
W_{2}^{2}(q_{Y^{n}},p_{Y}^{n}) & \le\mathbb{E}_{\hat{p}}[\|Y^{\prime n}-Y^{n}\|^{2}]\nonumber\\
 & =ab\mathbb{E}_{p}[\|Y^{\prime n}-Y^{n}\|^{2}]+\int_{B_{r}}\int_{B_{r}^{c}}\|y^{\prime n}-y^{n}\|^{2}\mathrm{d}q_{Y^{n}}(y^{\prime n})\mathrm{d}p_{Y}^{n}(y^{n})\nonumber\\
 & \quad+\int_{B_{r}^{c}}\int_{B_{r}}\|y^{\prime n}-y^{n}\|^{2}\mathrm{d}q_{Y^{n}}(y^{\prime n})\mathrm{d}p_{Y}^{n}(y^{n})+\int_{B_{r}^{c}}\int_{B_{r}^{c}}\|y^{\prime n}-y^{n}\|^{2}\mathrm{d}q_{Y^{n}}(y^{\prime n})\mathrm{d}p_{Y}^{n}(y^{n})\nonumber\\
 & \le4abr^{2}\mathbb{P}_{p}(Y^{\prime n}\neq Y^{n})+2b\int_{B_{r}^{c}}\|y^{\prime n}\|^{2}\mathrm{d}q_{Y^{n}}(y^{\prime n})+2(1-a)\int_{B_{r}}\|y^{n}\|^{2}\mathrm{d}p_{Y}^{n}(y^{n})\nonumber\\
 & \quad+2(1-b)\int_{B_{r}}\|y^{\prime n}\|^{2}\mathrm{d}q_{Y^{n}}(y^{\prime n})+2a\int_{B_{r}^{c}}\|y^{n}\|^{2}\mathrm{d}p_{Y}^{n}(y^{n})\nonumber\\
 & \quad+2(1-b)\int_{B_{r}^{c}}\|y^{\prime n}\|^{2}\mathrm{d}q_{Y^{n}}(y^{\prime n})+2(1-a)\int_{B_{r}^{c}}\|y^{n}\|^{2}\mathrm{d}p_{Y}^{n}(y^{n})\nonumber\\
 & \leq 4abr^{2}\|q_{Y^{n}}(\cdot|B_{r})-p_{Y}^{n}(\cdot|B_{r})\|_{\mathrm{TV}}+2\int_{B_{r}^{c}}\|y^{\prime n}\|^{2}\mathrm{d}q_{Y^{n}}(y^{\prime n})+2\int_{B_{r}^{c}}\|y^{n}\|^{2}\mathrm{d}p_{Y}^{n}(y^{n})\nonumber\\
 & \quad+2(1-b)\mathbb{E}_{q_{Y^{n}}}[\|Y^{n}\|^{2}]+2(1-a)\mathbb{E}_{p_{Y}^{n}}[\|Y^{n}\|^{2}]. \label{eq:-7}
\end{align}

We now establish the following inequality:
\begin{align}
\|q_{Y^{n}}-p_{Y}^{n}\|_{\mathrm{TV}}\ge (a\wedge b)\|q_{Y^{n}}(\cdot|B_{r})-p_{Y}^{n}(\cdot|B_{r})\|_{\mathrm{TV}}.\label{eq:TV}
\end{align}
There is no loss of generality in assuming $a\ge b$. Let $A\subseteq B_r$ be a set such that $q_{Y^n}(A|B_r)-p^n_{Y}(A|B_r)=\|q_{Y^{n}}(\cdot|B_{r})-p_{Y}^{n}(\cdot|B_{r})\|_{\mathrm{TV}}$.
It can be verified that
\begin{align}
\|q_{Y^{n}}-p_{Y}^{n}\|_{\mathrm{TV}} & =\max_{E}q_{Y^{n}}(E)-p_{Y}^{n}(E)\nonumber\\
 & \ge q_{Y^{n}}(A)-p_{Y}^{n}(A)\nonumber\\
 & =aq_{Y^{n}}(A|B_{r})-bp_{Y}^{n}(A|B_{r})\nonumber\\
 & \ge bq_{Y^{n}}(A|B_{r})-bp_{Y}^{n}(A|B_{r})\nonumber\\
 & =b\|q_{Y^{n}}(\cdot|B_{r})-p_{Y}^{n}(\cdot|B_{r})\|_{\mathrm{TV}},
\end{align}
which proves \eqref{eq:TV}.


Note that \eqref{eq:TV}, in conjunction with the fact $ab\leq a\wedge b$, implies 
\begin{align}
\|q_{Y^{n}}-p_{Y}^{n}\|_{\mathrm{TV}}\ge ab\|q_{Y^{n}}(\cdot|B_{r})-p_{Y}^{n}(\cdot|B_{r})\|_{\mathrm{TV}}.\label{eq:TVab}
\end{align}
Substituting \eqref{eq:TVab} into \eqref{eq:-7} yields 
\begin{align}
W_{2}^{2}(q_{Y^{n}},p_{Y}^{n}) & \le4r^{2}\|q_{Y^{n}}-p_{Y}^{n}\|_{\mathrm{TV}}+2\int_{B_{r}^{c}}\|y^{\prime n}\|^{2}\mathrm{d}q_{Y^{n}}(y^{\prime n})+2\int_{B_{r}^{c}}\|y^{n}\|^{2}\mathrm{d}p_{Y}^{n}(y^{n})\nonumber \\
 & \quad+2(1-b)\mathbb{E}_{q_{Y^{n}}}[\|Y^{n}\|^{2}]+2(1-a)\mathbb{E}_{p_{Y}^{n}}[\|Y^{n}\|^{2}].\label{eq:-3-2}
\end{align}
Taking the expectation over $\mathcal{C}$
on both sides of \eqref{eq:-3-2}, we have
\begin{align}
\mathbb{E}_{\mathcal{C}}[W_{2}^{2}(p_{Y^{n}|\mathcal{C}},p_{Y}^{n})]  &\le4r^{2}\mathbb{E}_{\mathcal{C}}[\|p_{Y^{n}|\mathcal{C}}-p_{Y}^{n}\|_{\mathrm{TV}}]+4\int_{B_{r}^{c}}\|y^{n}\|^{2}\mathrm{d}p_{Y}^{n}(y^{n})+4(1-b)\mathbb{E}_{p_{Y}^{n}}[\|Y^{n}\|^{2}]\nonumber\\
&=4r^{2}\mathbb{E}_{\mathcal{C}}[\|p_{Y^{n}|\mathcal{C}}-p_{Y}^{n}\|_{\mathrm{TV}}]+4\int_{B^c_{r}}\|y^{n}\|^{2}\mathrm{d}p_{Y}^{n}(y^{n})+4p_{Y}^{n}(B^c_{r})n\mathbb{E}[\|Y\|^{2}].\label{eq:-1-2}
\end{align}
Set $r=\sqrt{nN}$, where $N$ is a constant satisfying $N>\mathbb{E}[\|Y\|^{2}]$.
By the law of large numbers and the finiteness of $\mathbb{E}[\|Y\|^2]$,
\begin{align}
&\lim\limits_{n\rightarrow\infty}p_{Y}^{n}(B^c_{r})=0,\label{eq:law1}\\
&\lim\limits_{n\rightarrow\infty}\frac{1}{n}\int_{B^c_{r}}\|y^{n}\|^{2}\mathrm{d}p_{Y}^{n}(y^{n})=0.\label{eq:law2}
\end{align} 
Moreover, by the soft-covering lemma \cite{Hayashi06,Cuff13},
\begin{align}
\mathbb{E}_{\mathcal{C}}[\|p_{Y^{n}|\mathcal{C}}-p_{Y}^{n}\|_{\mathrm{TV}}]\to0\label{eq:law3}
\end{align}
exponentially fast as $n\to\infty$. Substituting \eqref{eq:law1}--\eqref{eq:law3} into
\eqref{eq:-1-2}  proves Lemma \ref{lem:soft-covering}.

Lemma \ref{lem:soft-covering} can be enhanced to guarantee exponential convergence, provided that  $p_{XY}$ satisfies a stronger condition.

\begin{lemma}
 Assume\footnote{It follows by Jensen's inequality that $\frac{1}{s}\ln(\mathbb{E}[e^{s\|Y\|^2}])\geq \mathbb{E}[\|Y\|^2]$. Therefore, $\mathbb{E}[e^{s\|Y\|^2}]<\infty$ implies $\mathbb{E}[\|Y\|^2]<\infty$.} that $p_{XY}$ satisfies $\mathbb{E}[e^{s\|Y\|^2}]<\infty$  for some $s>0$. If $R>I(X;Y)$, then $\mathbb{E}_{\mathcal{C}}[W_{2}^{2}(p_{Y^{n}|\mathcal{C}},p_{Y}^{n})]\to0$
exponentially fast as $n\to\infty$. 
\end{lemma}
\begin{IEEEproof}
We follow steps similar to those in the proof of Lemma  \ref{lem:soft-covering},
but instead set $r=\sqrt{nN}$ with $N>\frac{1}{s}\ln(\mathbb{E}[e^{s\|Y\|^2}])$.
By the Chernoff bound,
\begin{align} 
	p_{Y}^{n}(B^c_{r})\le e^{-ns\left(N-\frac{1}{s}\ln\mathbb{E}[e^{s\|Y\|^2}]\right)}\to0\label{eq:yield1}
\end{align}
exponentially fast as $n\to\infty$. Moreover, 
\begin{align}
\int_{B_{r}^{c}}\|y^{n}\|^{2}\mathrm{d}p_{Y}^{n}(y^{n}) & =\int_{0}^{\infty}p_{Y}^{n}(\|Y^{n}\|^{2}>(r^2\vee z))\mathrm{d}z\nonumber\\
 & =r^2p^n_Y(B^c_r)+\int_{r^{2}}^{\infty}p_{Y}^{n}(\|Y^{n}\|^{2}>z)\mathrm{d}z\nonumber\\
 & \le nNe^{-ns\left(N-\frac{1}{s}\ln\mathbb{E}[e^{s\|Y\|^2}]\right)}+\int_{nN}^{\infty}\mathbb{E}_{p_{Y}^{n}}[e^{s\|Y^{n}\|^{2}-sz}]\mathrm{d}z\nonumber\\
 & =nNe^{-ns\left(N-\frac{1}{s}\ln\mathbb{E}[e^{s\|Y\|^2}]\right)}+\frac{1}{s}(\mathbb{E}[e^{s\|Y\|^2}])^{n}e^{-nsN}\nonumber\\
 & =e^{-ns\left(N-\frac{1}{s}\ln\mathbb{E}[e^{s\|Y\|^2}]-\frac{1}{ns}\ln(nN)\right)}+\frac{1}{s}e^{-ns\left(N-\frac{1}{s}\ln\mathbb{E}[e^{s\|Y\|^2}]\right)}\nonumber\\
 &\rightarrow 0\label{eq:yield2}
\end{align}
 exponentially fast as $n\to\infty$.
Substituting \eqref{eq:law3}, \eqref{eq:yield1}, and \eqref{eq:yield2}  into
\eqref{eq:-1-2} yields the desired claim. 
\end{IEEEproof}

\section{Proof of Theorem \ref{thm:Gaussianvector}}\label{app:Gaussianvector}

First we shall show that 
\begin{align}
	D(R,C,P)\leq\sum\limits_{\ell=1}^L\gamma_{\ell} 2^{-2r_{\ell}}+\left[\left(\sqrt{\sum\limits_{\ell=1}^L\gamma_{\ell}(2-2^{-2r_{\ell}}-2\psi(r_{\ell},r'_{\ell}))}-\sqrt{P}\right)_+\right]^2\label{eq:any}
\end{align}
for any $(r_1,r_2,\ldots,r_L),(r'_1,r'_2,\ldots,r'_L)\in\mathbb{R}^L_+$ satisfying \eqref{eq:sum1} and \eqref{eq:sum2}. Let $\tilde{X}:=(\tilde{X}_1,\tilde{X}_2,\ldots,\tilde{X}_L)^T\sim\mathcal{N}(0,\tilde{\Gamma})$, where $\tilde{\Gamma}:=\mathrm{diag}(\gamma_1(1-2^{-2r_1}),\gamma_2(1-2^{-2r_2}),\ldots,\gamma_L(1-2^{-2r_L}))$. Moreover, let $\mu$  and $\nu$ be two couplings of $p_X$ and $p_{\tilde{X}}$ such that 
\begin{enumerate}
	\item  $(X_1,\tilde{X}_1),(X_2,\tilde{X}_2),\ldots,(X_L,\tilde{X}_L)$ are mutually independent under both $\mu$ and $\nu$, 
	
	\item  for $\ell=1,2,\ldots,L$, the covariance matrix of $(X_{\ell},\tilde{X}_{\ell})$ is given by
	\begin{align}
		\left(\begin{matrix}
			\gamma_{\ell} & \gamma_{\ell}(1-2^{-2r_{\ell}})\\
			\gamma_{\ell}(1-2^{-2r_{\ell}}) & \gamma_{\ell}(1-2^{-2r_{\ell}})
		\end{matrix}\right)\quad \mbox{under }\mu,
	\end{align}
	and by
	\begin{align}
		\left(\begin{matrix}
			\gamma_{\ell} & \gamma_{\ell}\psi(r_{\ell},r'_{\ell})\\
			\gamma_{\ell}\psi(r_{\ell},r'_{\ell}) & \gamma_{\ell}(1-2^{-2r_{\ell}})
		\end{matrix}\right)\quad \mbox{under }\nu.
	\end{align}
\end{enumerate}
It can be verified that
\begin{align}
	&\mathbb{E}_{\mu}[X|\tilde{X}]=\tilde{X}\quad\mu-\mbox{a.s.},\\
	&I_{\mu}(X;\tilde{X})=\sum\limits_{\ell=1}^Lr_{\ell}\leq R,\\
	&I_{\nu}(X;\tilde{X})=\sum\limits_{\ell=1}^Lr'_{\ell}\leq R+C,\\
	&\mathbb{E}_{\mu}[\|X-\tilde{X}\|^2]=\sum\limits_{\ell}^L\gamma_{\ell} 2^{-2r_{\ell}},\\
	&\mathbb{E}_{\nu}[\|X-\tilde{X}\|^2]=\sum\limits_{\ell=1}^L\gamma_{\ell}(2-2^{-2r_{\ell}}-2\psi(r_{\ell},r'_{\ell})).
\end{align}
This proves \eqref{eq:any}.

Next we proceed to show that
\begin{align}
	D(R,C,P)=\sum\limits_{\ell=1}^L\gamma_{\ell} 2^{-2r_{\ell}}+\left[\left(\sqrt{\sum\limits_{\ell=1}^L\gamma_{\ell}(2-2^{-2r_{\ell}}-2\psi(r_{\ell},r'_{\ell}))}-\sqrt{P}\right)_+\right]^2\label{eq:some}
\end{align}
for some $(r_1,r_2,\ldots,r_L),(r'_1,r'_2,\ldots,r'_L)\in\mathbb{R}^L_+$ satisfying \eqref{eq:sum1} and \eqref{eq:sum2}. Given any $p_{\tilde{X}}\in\mathcal{P}(\mathbb{R}^L)$ and $\mu,\nu\in\Pi(p_X,p_{\tilde{X}})$  with
\begin{align}
	&\mathbb{E}_{\mu}[X|\tilde{X}]=\tilde{X}\quad\mu-\mbox{a.s.},\label{eq:asv}\\
	&I_{\mu}(X;\tilde{X})\leq R,\label{eq:Rconstrainv}\\
	&I_{\nu}(X;\tilde{X})\leq R+C,\label{eq:R+Cconstraintv}
\end{align}
let $\tilde{\gamma}_{\ell}:=\mathbb{E}[\tilde{X}^2_{\ell}]$ and $\theta:=\mathbb{E}_{\nu}[X_{\ell}\tilde{X}_{\ell}]$, $\ell=1,2,\ldots,L$. It can be shown by following the corresponding steps in the scalar case that for $\ell=1,2,\ldots,L$,
\begin{align}
	&I_{\mu}(X_{\ell};\tilde{X}_{\ell})\geq r_{\ell}:=\frac{1}{2}\log\left(\frac{\gamma_{\ell}}{\gamma_{\ell}-\tilde{\gamma}_{\ell}}\right),\\
	&I_{\nu}(X_{\ell};\tilde{X}_{\ell})\geq r'_{\ell}:=\frac{1}{2}\log\left(\frac{\gamma_{\ell}\tilde{\gamma}_{\ell}}{\gamma_{\ell}\tilde{\gamma}_{\ell}-\theta^2_{\ell}}\right),\\
	&\mathbb{E}_{\mu}[(X_{\ell}-\tilde{X}_{\ell})^2]=\gamma_{\ell}2^{-2r_{\ell}},\\
	&\mathbb{E}_{\nu}[(X_{\ell}-\tilde{X}_{\ell})^2]=\gamma_{\ell}(2-2^{-2r_{\ell}}-2\psi(r_{\ell},r'_{\ell})).
\end{align} 
Therefore, we have
\begin{align}
	&\sum\limits_{\ell=1}^Lr_{\ell}\leq\sum\limits_{\ell=1}^LI_{\mu}(X_{\ell};\tilde{X}_{\ell})\leq I_{\mu}(X;\tilde{X})\leq R,\\
	&\sum\limits_{\ell=1}^Lr'_{\ell}\leq\sum\limits_{\ell=1}^LI_{\nu}(X_{\ell};\tilde{X}_{\ell})\leq I_{\nu}(X;\tilde{X})\leq R+C,\\
	&	\mathbb{E}_{\mu}[\|X-\tilde{X}\|^2]=\sum\limits_{\ell=1}^L\gamma_{\ell}2^{-2r_{\ell}},\\
	&\mathbb{E}_{\nu}[\|X-\tilde{X}\|^2]=\sum\limits_{\ell=1}^L\gamma_{\ell}(2-2^{-2r_{\ell}}-2\psi(r_{\ell},r'_{\ell})).
\end{align}
This proves \eqref{eq:some}.

In view of \eqref{eq:any} and \eqref{eq:some}, the proof is complete by invoking Theorem \ref{thm:main}.




\ifCLASSOPTIONcaptionsoff
  \newpage
\fi

\end{document}